\pdfoutput=1
\documentclass[preprint,journal]{vgtc}       

\ifpdf
  \pdfoutput=1\relax                   
  \usepackage{graphicx}                
  \DeclareGraphicsExtensions{.pdf,.png,.jpg,.jpeg} 
\else
  \usepackage{graphicx}                
  \DeclareGraphicsExtensions{.eps}     
\fi%

\graphicspath{{figures/}{pictures/}{images/}{./}} 

\usepackage{microtype}                 
\PassOptionsToPackage{warn}{textcomp}  
\usepackage{textcomp}                  
\usepackage{mathptmx}                  
\usepackage{times}                     
\usepackage{cite}                      
\usepackage{tabu}                      
\usepackage{booktabs}                  

\usepackage{wrapfig, lipsum}
\usepackage{fontawesome}
\usepackage{amsfonts,amssymb}
\usepackage{xr}

\usepackage{amsmath}

\onlineid{1123}

\vgtccategory{Research}
\vgtcpapertype{Analytics \& Decisions}

\title{Towards Visual Explainable Active Learning \\ for Zero-Shot Classification}

\author{Shichao Jia, Zeyu Li, Nuo Chen, \textit{Student Member, IEEE}, and Jiawan Zhang, \textit{Senior Member, IEEE}}
\authorfooter{
\item
 S. Jia, Z. Li, N. Chen and J. Zhang are with College of Intelligence and Computing, Tianjin University. E-mail: \{jsc\_se, lzytianda, nicole\_0420, jwzhang\}@tju.edu.cn. J. Zhang is the corresponding author.
 \item
 J. Zhang is also with Tianjin cultural heritage conservation and inheritance engineering technology center and Key Research Center for Surface Monitoring and Analysis of Relics, State Administration of Cultural Heritage, China.
}

\shortauthortitle{Biv \MakeLowercase{\textit{Jia et al.}}: Towards Visual Explainable Active Learning for Zero-Shot Classification}

\abstract{Zero-shot classification is a promising paradigm to solve an applicable problem when the training classes and test classes are disjoint. Achieving this usually needs experts to externalize their domain knowledge by manually specifying a class-attribute matrix to define which classes have which attributes. Designing a suitable class-attribute matrix is the key to the subsequent procedure, but this design process is tedious and trial-and-error with no guidance. This paper proposes a visual explainable active learning approach with its design and implementation called semantic navigator to solve the above problems. This approach promotes human-AI teaming with four actions (ask, explain, recommend, respond) in each interaction loop. The machine asks contrastive questions to guide humans in the thinking process of attributes. A novel visualization called semantic map explains the current status of the machine. Therefore analysts can better understand why the machine misclassifies objects. Moreover, the machine recommends the labels of classes for each attribute to ease the labeling burden. Finally, humans can steer the model by modifying the labels interactively, and the machine adjusts its recommendations. The visual explainable active learning approach improves humans' efficiency of building zero-shot classification models interactively, compared with the method without guidance. We justify our results with user studies using the standard benchmarks for zero-shot classification. %
} 


\CCScatlist{ 
 \CCScat{K.6.1}{Management of Computing and Information Systems}%
{Project and People Management}{Life Cycle};
 \CCScat{K.7.m}{The Computing Profession}{Miscellaneous}{Ethics}
}

\teaser{
  \centering
  \includegraphics[width=\linewidth]{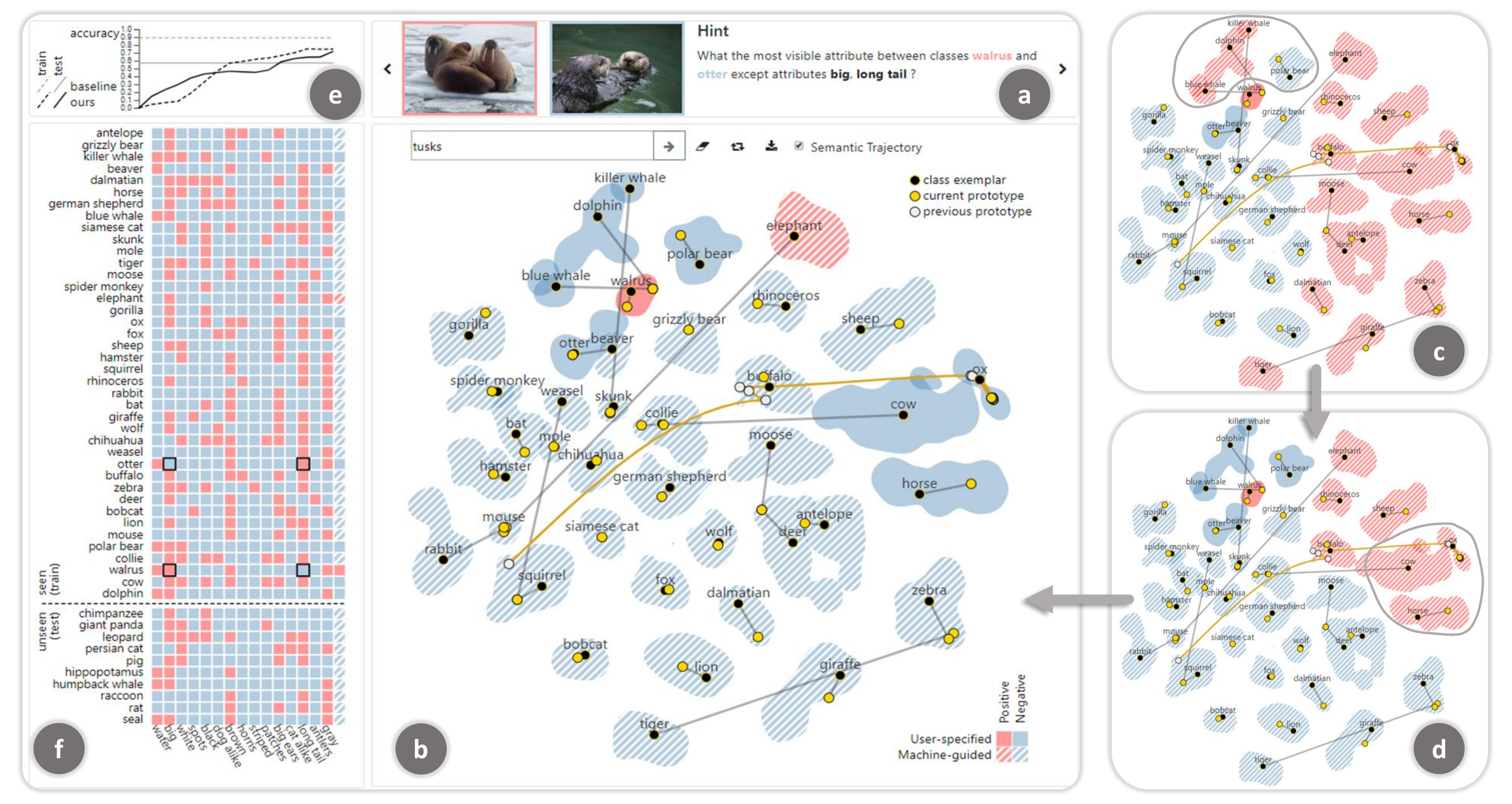}
  \caption{Semantic navigator is a mixed-initiative visual analytics system for zero-shot classification. (a) The machine asks contrastive questions to guide analysts to come up with new attributes. (b) The semantic map explains the machine's status and presents the label recommendations (striped contours). Analysts select partial classes as positive (solid red contours) or negative (solid blue contours) to adjust the label recommendations ((c) and (d)). (e) The line chart monitors the training accuracy for seen classes and testing accuracy for unseen classes. (f) The class-attribute matrix is built interactively via collaboration between analysts and the machine. }
  \label{fig:teaser}
}

\vgtcinsertpkg

\begin{document}


\firstsection{Introduction} 

\maketitle

Zero-shot classification \cite{xian2019zero-shot} is a dominant and promising learning paradigm, aiming to recognize instances that are unseen during training. Compared with traditional supervised classification, which needs enormous labeled training data and can only classify instances covered by seen classes, zero-shot classification is more applicable and more like the human reasoning process. Humans can recognize unseen objects once we provide the description of objects. For example, we can teach children to recognize zebras by describing that zebras look like horses with black and white stripes. Zero-shot classification is based on this idea, and due to its practical value, it gains lots of attention in recent machine learning and computer vision studies \cite{wang2019survey}.

\begin{figure}
 \centering 
 \includegraphics[width=1.0\linewidth]{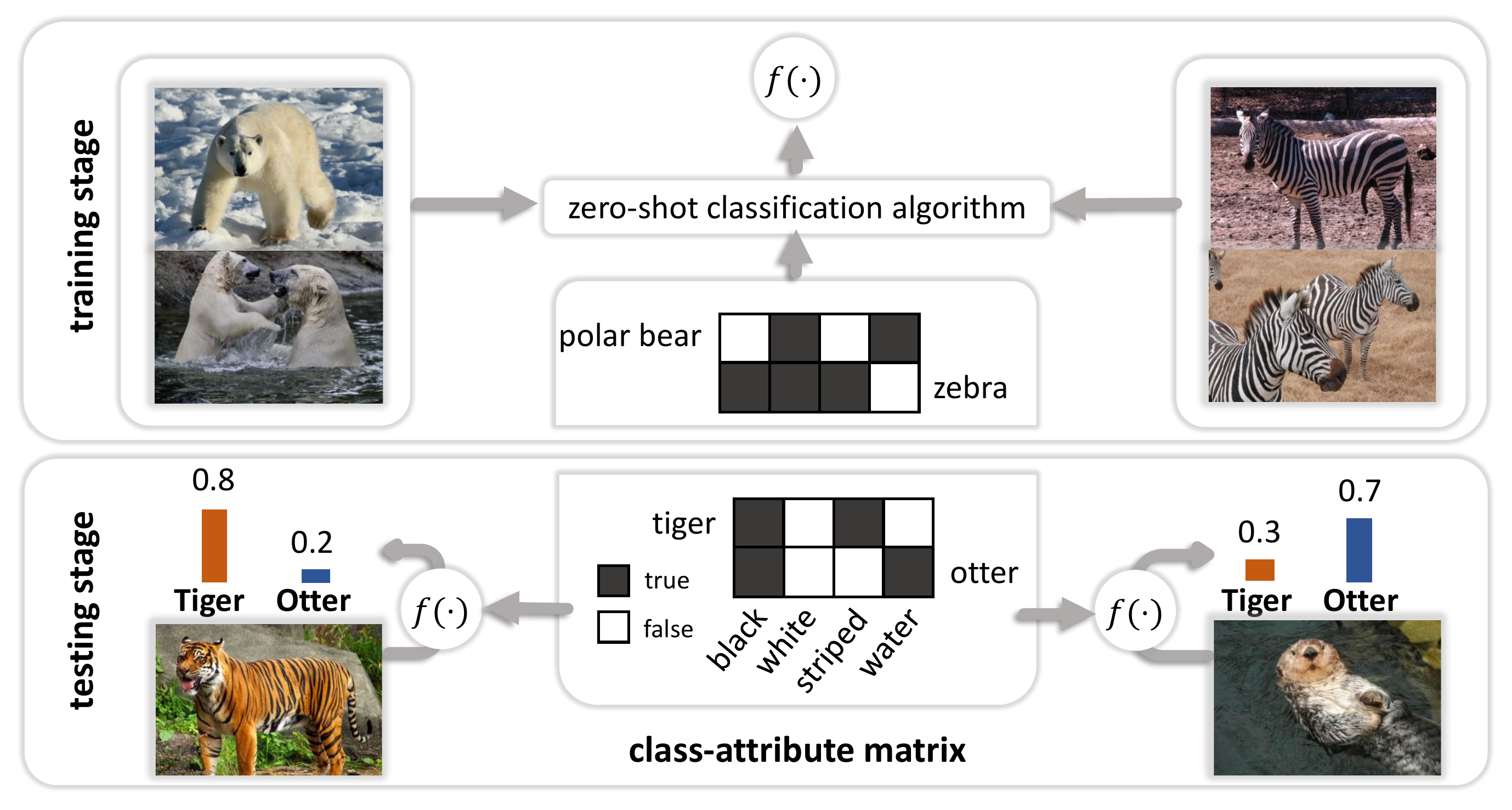}
 \caption{The basic framework of zero-shot classification. Leveraging the class-attribute matrix using visual attributes, zero-shot classifier ($f(\cdot)$) can recognize tiger and otter even though the training data does cover these unseen classes. The class-attribute matrix is central to the zero-shot classification. However, it is very challenging to design a suitable class-attribute matrix manually without guidance. }
 \label{fig:zero-shot}
\end{figure}

One critical process to achieve zero-shot classification is building a class-attribute matrix (matrix with rows as classes and columns as attributes, see Fig.\ref{fig:zero-shot}). The attributes are human interpretable semantic words designed by experts, such as big eyes, black hair, etc. We can distinguish classes (including seen and unseen classes) by describing whether or how likely they have the attributes. Once the machine learns the concepts of attributes from seen classes, it can recognize objects from unseen classes leveraging the description using attributes.

However, constructing the class-attribute matrix is very challenging. Firstly, this process is hard to be automated with interpretability preserved. Designing attributes usually involves manually picking descriptive words for the images under consideration by domain specialists \cite{farhadi2009describing, lampert2009learning}. Yu et al. \cite{yu2013designing} try to avoid human labor to build the class-attribute matrix automatically. However, the attributes designed by this approach have no concise names and may not be semantically interpreted. Secondly, the attributes used in the zero-shot classification are selective. Not all attributes are necessary and helpful for zero-shot classification, and some are even harmful and lead to wrong predictions \cite{guo2018zero-shot}. Thirdly, once the analyst \footnote{In this paper, the analysts (or end-users) are data scientists and domain experts who want to build zero-shot classifiers by injecting their domain knowledge with limited machine learning knowledge.} comes up with new attributes, he/she needs to go through the tedious and painful process to label every class. In summary, up to now, it can be tricky to design the class-attribute matrix for zero-shot classification using solely machine-centric processes or solely human-centric processes.


One promising approach to the problem is teaming the machine and analysts with the machine guiding the attribute design process. Approaches to alleviating the labeling burden with the machine querying labels from humans, aka active learning \cite{settles2009active}, have been widely used in traditional supervised learning. While in this paper, instead of asking for labels of instances, the machine seeks attributes to differentiate classes.  However, active learning merely recognizes humans as labeling machines so that analysts can not steer the process. It has shown that combining visual interactive labeling with active learning in a mixed-initiative way can accelerate the labeling process \cite{bernard2017comparing, bernard2018towards}. Meanwhile, explainability is a crucial factor in promoting human-AI teaming \cite{miller2019explanation} as humans need to understand the current state of the machine. Considering these, we propose a visual explainable active learning approach to zero-shot classification with four key actions (ask, explain, recommend and respond): machine asks questions to guide analysts to come up with new attributes; visualization explains the current state of the machine; machine recommends labels of attributes for each class; analysts provide feedback to the recommendations and steer the process.

Achieving the above human-AI teaming tasks needs to facilitate a shared mental model between the analyst and the machine \cite{zhang2021ideal, kaur2019building, hong2018coordinating}. Without understanding the state of the machine, the analyst will not trust the machine \cite{chatzimparmpas2020state} and can not know how to steer the process. Similarly, without inferring how the analyst reasons the task, the machine can not aid the design process. Specifically, current deep neural networks recognize objects utilizing enigmatic features that humans hardly understand.  Contrastively, humans, describe classes leveraging high-level interpretable attributes, which is not the logic to distinguish objects for the machine. This semantic gap between the machine and the analyst seriously hinders the effectiveness of human-AI teaming.

To solve this technical challenge, we bridge the two spaces of features and attributes into one shared space facilitating mutual coordination. We then design a semantic map to visualize this space as the main interface for cooperative attribute design. This visualization enables analysts to reason the misclassification of the zero-shot model and inspire them to come up with new attributes. Based on this, we further design and implement a mixed-initiative visual analytics system called semantic navigator to facilitate other teaming tasks. The semantic navigator guides analysts with contrastive questions to elicit attributes by comparing two class exemplars. Contrastive questions require analysts to focus on the differences between classes, which is easier than the method to let humans come up with new attributes directly. Meanwhile, we leverage semisupervised support vector machines to provide label recommendations and enable analysts to interactively provide feedback to the recommendations.

In summary, our technical contributes are as follows:

\begin{itemize}
  \item We propose a human-AI teaming approach called visual explainable active learning to elicit human knowledge and enable analysts to steer models at the same time.

  \item We design and implement a visual analytics system called semantic navigator that guides analysts to interactively build zero-shot classification models.

  \item We evaluate our technique with case studies and user studies. We find the semantic navigator can improve the efficiency of analysts and accuracy of zero-shot models compared with the method without guidance.
\end{itemize}

\section{Related Work}

In this section,  we describe prior work relevant to our study, including methods of designing visual attributes (Sec. \ref{design_via}), interactive classification (Sec. \ref{iML}) and guidance in  visual analytics (Sec. \ref{mixed_va}).

\subsection{Designing Visual Attributes} \label{design_via}

Traditionally, visual attributes are designed by manually picking a set of words that are descriptive and distinguishable for the images \cite{farhadi2009describing, lampert2009learning}. Therefore, this process needs special expertise and great effort for domain experts, which is expensive for laypersons. To alleviate the burdens, Berg et al. \cite{berg2010automatic} propose an automatic approach to tag images with visual attributes by mining text and image data sampled from the Internet. However, the discovered attributes from text-mining can contain irrelevant semantics for the classification task at hand or may not be separable in the visual feature space. Considering this, Parikh et al. \cite{parikh2011interactively} propose to build nameable and discriminative attributes with human-in-the-loop. Based on that, Duan et al. \cite{duan2012discovering} propose a method to discover localized attributes for fine-grained recognition. Yu et al. \cite{yu2013designing} propose an automatic method to design discriminative category-level attributes. However, the learned attributes will not have concise semantics as the manually specified attributes. Moreover, for unseen classes, this approach needs humans to model the similarity between seen classes and unseen classes, which is more difficult and unintuitive than directly specifying visual attributes of unseen classes. Considering interpretability is inevitable in making human and machine communication and trust, we step back to think about improving human efficiency when designing visual attributes with human-in-the-loop.

\subsection{Interactive Classification} \label{iML}

Interactive machine learning \cite{fails2003interactive}, or machine teaching \cite{simard2017machine}, places a human-centered perspective on the building process of machine learning models. This paradigm leverages end-user involvement to enable rapid, focused, and incremental interaction cycles with various applications \cite{jiang2019recent}. Visualization as the interface between humans and machines plays a crucial role in seamlessly fitting analytics into existing interactive process \cite{endert2014human}. We mainly discuss methods of interactive building classification models.


Labeling data instances is an important task for interactive classification. Active learning \cite{settles2009active} takes the machine-centric approach to ease the labeling burden by querying labels from users. In contrast, visual-interactive labeling \cite{seifert2010user} takes the human-centric approach leveraging humans' ability to identify patterns with interactive visualization. Bernard et al. \cite{bernard2017comparing, bernard2018vial} systematically compare the performance of the two paradigms and summarize different user strategies for visual interactive labeling. Heimerl et al. \cite{heimerl2012visual} compare different combinations of active learning and visual interactive learning for document retrieval. Similarly, H\"{o}ferlin et al. \cite{hoferlin2012inter} combine the two approaches for video visual analytics, and Sun et al. \cite{sun2017label} present Label-and-Learn system to help analysts understand the classifier's behavior. Jose Gustavo et al. \cite{paiva2014approach} propose an incremental visual classification method that enables users to steer the classification process by annotating items inside the similarity tree of data. Liu et al. \cite{liu2018interactive} propose a visual analytics method to improve the crowdsourced annotations while improving the classification model. Xiang et al. \cite{xiang2019interactive} solve a similar problem with a different approach called DataDebugger. Felix et al. \cite{felix2018exploratory} propose exploratory labeling to facilitate label ideation, specification, and refinement with machine-driven recommendations. Our research expands the research space considering attribute ideation guided by the machine and label recommendations with human feedback for zero-shot classification.



Sahoo et al. \cite{Sahoo2020VisuallyAA} step first to study interactive zero-shot classification for diagnosing and steering its procedure. However, this process is under the assumption that the class-attribute matrix is provided. In contrast, our research focuses on a more fundamental problem that how to facilitate the constructing process of the class-attribute matrix. Analysts build classification models by describing classes using visual attributes via collaboration with the machine. Another advantage of our method is that the whole process is naturally interpretable when visual features extracted from the deep neural networks are involved. Our technique contributes to another way of interpretable machine learning \cite{hohman2018visual, choo2018visual}. Previous works either open the black boxes \cite{liu2016towards, strobelt2017lstmvis, liu2017visual, ming2017understanding, liu2017analyzing, wang2018dqnviz} which need analysts to have machine learning knowledge or explain the black boxes using surrogate models \cite{ming2018rulematrix, jia2020visualizing}, which adds another layer of uncertainty.


\subsection{Guidance} \label{mixed_va}

Guidance is a mixed-initiative process to assist analysts with a set of visual means to close the knowledge gap via the collaboration of the machine. Ceneda et al. \cite{ceneda2019review} present a systematic review of guidance approaches in visual analytics based on the dimensions of the knowledge generation process. We only discuss the most relevant methods that guide model construction and parameter refinement processes.

\begin{figure*}
 \centering 
 \includegraphics[width=1.0\textwidth]{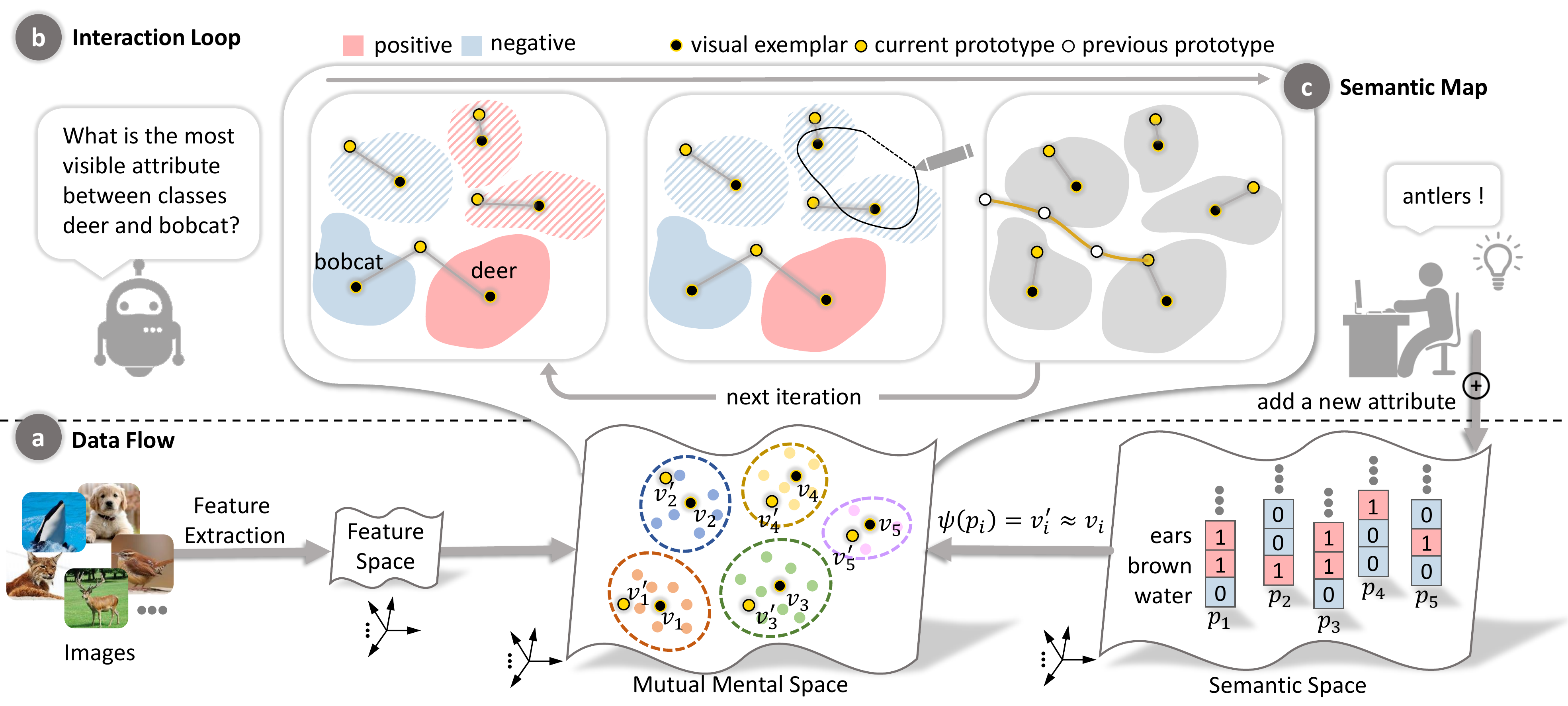}
 \caption{Visual explainable active learning for zero-shot classification: (a) The zero-shot classifier first leverages features usually extracted from a deep neural network (DNN) as input. The mutual mental space is then constructed by mapping both the feature space and semantic space into one shared space. The zero-shot classifier learns a mapping ($\psi$) to predict the visual exemplars (centers of class clusters in the mutual mental space, represented as $v_i$) using the class prototypes ($p_1$-$p_5$) specified by the analyst in the semantic space. The predicted exemplars are denoted as $v_i'$. In this paper, the feature space is in 2048-dimensional space while the mutual mental space is in 500-dimensional space. (b) The semantic navigator closes each interaction loop with four key actions (ask, explain, recommend and respond). (c) The semantic map visualizes data distributions as contours, and each pair of the visual exemplar and prototype linked by a gray line. It supports explanations of the machine's status, label recommendations, and user feedback by interactively correcting annotations to ease the burden of specifying attributes. }
 \label{fig:pipeline}
\end{figure*}

Dis-Function \cite{brown2012dis-function} enables analysts to directly manipulate the data points to express the similarity by the proximity, and machines transform the interactions into the weights of distance functions. This idea has inspired a new interaction paradigm called semantic interaction \cite{endert2014semantic} to infer analysts' interests by manipulating visual items directly. For example, Endert et al. \cite{endert2012semantic} propose requireSPIRE to steer text models by spatially clustering data points. Moreover, Dowlinget et al. extend semantic interaction to a bidirectional pipeline to infer both the observation and attribute weights \cite{dowling2019sirius}. Sacha et al. \cite{sacha2017somflow} propose a guided visual analytics system called SOMFlow for exploratory cluster analysis using self-organizing maps. Specifically, SOMFlow provides a ranked list of recommendations and visual cues to guide analysts. Cavallo et al. propose Clustrophile 2 \cite{cavallo2018clustrophile} for guided visual clustering analysis. This system guides analysts to explore the large clustering space and finetune the clustering parameters. Other techniques \cite{o2015designscape, bouali2016vizassist, wongsuphasawat2015voyager, vartak2015seedb} aim to provide suggestions to guide the visualization design or graphic design. Our work focuses on guided supervised classification for zero-shot learning by construcing the class-attribute matrix, which has not been explored yet.

\section{Human-AI Teaming for Zero-Shot Classification}

%
%
%
%
%
%

\subsection{Problem Statement}

This section briefly introduces the basic concepts of zero-shot classification and formally states the research problem, which is essential for subsequent discussion.

In zero-shot classification, the classes covered by training data are {\textbf{seen classes}, denoted as $C_s = \left\{c_i^s | i = 1,...,N_s \right\}$. Contrastively, classes for unlabeled testing instances are \textbf{unseen classes}, denoted as $C_u = \left\{c_i^u | i = 1,...,N_u\right\}$. The seen classes and unseen classes are disjoint, namely $C_s \cap \ C_u = \emptyset$. Both training and testing data come from a $D$ dimensional real space denoted as $\mathcal{X} \in \mathbb{R}^D$, namely \textbf{feature space}. We denote $D_{tr} = \left\{(x_i^{tr}, y_i^{tr}) \in \mathcal{X} \times C_s\right\}$ as the training dataset, in which $x_i^{tr}$ is the labeled instance in the feature space, and $y_i^{tr}$ is the corresponding class label in the seen classes. Similarly, we denote $D_{te} = \left\{ x_i^{te} \in \mathcal{X} \right\}$ as the testing dataset, in which $x_i^{te}$ is the unlabeled instance in the feature space. Given the training dataset $D_{tr}$, zero-shot classification aims to learn a classifier $f( \cdot ) : \mathcal{X} \rightarrow C_u$ that predicts the labels of testing dataset $D_{te}$.

As there are no instances of unseen classes during training, the machine needs some auxiliary information to transfer knowledge from the seen classes to unseen classes. The most widely used auxiliary information are \textbf{attributes}. Attributes are human understandable words to describe objects with meaningful characteristics. These attributes construct a space namely \textbf{semantic space} ($\mathcal{S} \in \mathbb{R}^M $ ($M < D$)), with each dimension being one attribute. Attributes can be binary ($0/1$) to symbolize their presence, or continuous to represent the degree of confidence. In this paper, we consider binary attributes as the initial step of the research. In the semantic space, each class is represented as a semantic vector namely \textbf{class prototype}. For example, in Fig. \ref{fig:zero-shot}, there are four visual attributes to describe classes, including black, white, striped, and water. The zebra prototype is specified as $[1, 1, 1, 0]$. We denote the prototypes of seen classes as $P_s = \left\{ p_i^s | p_i^s \in \mathcal{S} \right\}_{i=1}^{N_s}$, and those of unseen classes as $P_u = \left\{ p_i^u | p_i^u \in \mathcal{S} \right\}_{i=1}^{N_u}$. $D_{tr}$, $P_s$, $P_u$ are vital to obtain a zero-shot classifier $f(\cdot)$. Aligning all the class prototypes vertically, we get the \textbf{class-attribute matrix}, denoted as $S \in \left\{ 0, 1 \right\}^{ N \times M}$ ($N = N_s + N_u$). Our paper aims to build the binary class-attribute matrix interactively and train a good function $f(\cdot)$ as well.

\subsection{Visual Explainable Active Learning} \label{app_overview}


Analysts without systematic domain knowledge can not easily transform the implicit knowledge to the structural class-attribute matrix. Although humans make decisions using attributes every minute, this process is usually subconscious and difficult to be characterized. To tackle this issue, we propose visual explainable active learning for zero-shot classification with four actions (ask, explain, recommend, respond) to guide the design process.

\noindent\textbf{Ask.}
Considering dialogue is a natural way for human interactions, the machine can elicit the attributes by asking questions. A naive approach is to ask for attributes given pictures of specific categories directly. However, this approach is inefficient because direct questions are not well-defined and may elicit irrelevant attributes to the classification. Considering this, we adopt contrastive questions by letting analysts compare two representative images of different categories and discover attributes that differentiate the two classes. These questions are simple enough but powerful by leveraging humans' analogical reasoning \cite{gentner2003learning} to discover deeper structural characteristics. Notice that we do not constrain the forms of questions, as there are still limited studies on how different kinds of questions impact the elicitation process \cite{tian2017learning}.


\noindent\textbf{Explain.}
During the training process, visualization can play a critical role in enabling analysts to interpret and diagnose results from the machine.  One fundamental technical problem in this setting is that features from the machine and attributes elicited from the analysts lie in the two different spaces. The semantic gap between the two spaces hinders the effectiveness of human-AI teaming. It is vital to bridge the two spaces using visualization to enhance the explainability of the results. We solve this problem by mapping the feature space and semantic space into one shared space, and we name the new space as \textbf{mutual mental space}. To explain the zero-shot classification results in the mutual mental space, we leverage explainable case-based reasoning \cite{lamy2019explainable}. Numerous studies \cite{ming2019protosteer} have demonstrated that case-based reasoning involving various forms of matching and prototyping is fundamental to effective decision-making strategies. In this paper, we explain the zero-shot classifier results by visualizing the relationships between prototypes and corresponding desirable exemplars (details in Sec. \ref{layout}).

\noindent\textbf{Recommend and Respond.}
Meanwhile, eliciting a new attribute in every interaction loop is just one step towards building the class-attribute matrix. Labeling classes using the attributes is another burden. We enable the machine to recommend labels to fill the blanks in the class-attribute matrix. Moreover, analysts can respond to recommendations by interactively modifying class labels through visualization to close the interaction loop. We call the above approach containing four actions (ask, explain, recommend, respond) as visual active learning for zero-shot classification.

\section{Semantic Navigator} \label{orient_guidance}

\subsection{System Overview}

We instantiate the visual explainable active learning for zero-shot classification as a visual analytic approach called semantic navigator. We first discuss how we construct the mutual mental space as this is the technical fundamentals for visualization. Then we provide the overview of visualization views and interaction loop. Finally, we cover the training and testing process of the underlying zero-shot classifier, as its running process closely relates to visualization and interaction.

\noindent\textbf{Mutual Mental Space.} \label{p:training}
We use PCA to project these features into the mutual mental space. These features are usually extracted from a pre-trained deep neural network. PCA decorates the dimensions of features so that we can predict these dimensions independently rather than jointly. Besides, we apply PCA to gain computational benefits by reducing dimensionality, considering the need for real-time interaction. We denote the PCA projection matrix as $M_{PCA} \in \mathbb{R}^{ d \times D } (2 < d < D)$. Notice that we choose the dimensionality $d$ empirically in this paper, and the projected space using PCA is still in high dimensional space. For example, in Sec. 5, we set $d = 500$ when the feature space is in 2048-dimensional space. Besides, $M_{PCA}$ is computed over all the training data and not class-specific. In the mutual mental space, the data for each class usually form a cluster. We assume that each cluster center is our target semantic representation, namely \textbf{visual exemplars}. Then we map the prototypes into the new space by predicting the visual exemplars from the class prototypes.

\noindent\textbf{Visualization Views.}
The semantic map (Fig. \ref{fig:pipeline}(c)) is the critical visualization of the semantic navigator. It explains the status of the zero-shot model, visualizes the label recommendations, and enables feedback interaction from analysts. Each contour in the semantic map presents one class cluster. The interactive zero-shot classification targets to narrow the gap between each pair of the class prototype (yellow dot) and corresponding visual exemplar (black dot). The colors of the striped contours stand for the recommended labels of the current attribute. Besides, the semantic navigator contains three more views: a hint view (Fig. \ref{fig:teaser}(a)), a
matrix view (Fig. \ref{fig:teaser}(f)), and a line chart (Fig. \ref{fig:teaser}(e)). The hint view presents a ranked list of questions to elicit attributes from the analyst. Visualizing the class-attribute matrix enables the analyst to scrutinize the externalized knowledge. The line chart presents the training and testing accuracy and enables monitoring status of the current zero-shot classifier.

\noindent\textbf{Interaction Loop.}
The workflow closes the loop with four actions (ask, explain, recommend, respond), seen in Fig. \ref{fig:pipeline}(b). The machine asks contrastive questions (what is the most visible attribute between classes A and B except existing attributes) to guide analysts to come up with new attributes. Once the analyst comes up with a new attribute, he/she can interactively specify the positive or negative status of partial classes using lasso interaction. The machine recommends the rest classes to ease the labeling burden. After specifying a new attribute, the analyst can click the submit button. The semantic navigator will pop up a window to let the analyst specify unseen classes with the new attribute. After specification, the underlying model is trained on the new class-attribute matrix, and the visualization will be updated. A new iteration begins until the gaps are small enough, or human resources are exhausted.

The interaction loop challenges the current zero-shot classification models as they need to be computationally efficient to support real-time interaction. To this end, we choose one state-of-art model called EXEM \cite{changpinyo2017predicting} as its runtime depends only on the number of training classes, while most other methods can not. Notice that this model is not a must for our approach. Any model that builds a shared space between the feature space and the semantic space with efficient computation can be applied.

\noindent\textbf{Training and Testing.}
Specifically, the EXEM model learns a transformation function $\psi(\cdot)$ from semantic space to the mutual mental space for each class $i$ using $d$ support vector regressors (SVR), such that $\psi(p_i^s) \approx v_i^s$, where the $p_i^s (p_i^s \in P_s)$ is the seen class prototype, and $v_i^s \in \mathbb{R}^d$ is the visual exemplar for class $i$. Each SVR predicts each dimension of the visual exemplars from their corresponding class prototypes. Denote the set of visual exemplars for seen classes as $V_s = \left\{ v_i^s | v_i^s \in \mathbb{R}^d \right\}$. Now we can perform zero-shot classification using a nearest neighbor classifier, once we provide the prototypes of unseen classes $p_i^u (p_i^u \in P_u)$ by specifying their attributes. That is, the classifier outputs the label of the closest exemplar for each novel data point $x_u$, namely $\overline{y} = \arg\min_{i} dist (M_{PCA}x_u, \psi(p_i^u))$.

In the following sections, we discuss how to close the loop with four actions (ask, explain, recommend, and respond) supported by the semantic navigator.

\subsection{Question Generation} \label{hints}

One design principle for the hint view is that it should ask questions with upgrading difficulty as the number of attributes increases, as it fits humans' cognition. Otherwise, it will discourage the analyst from making progress. Therefore at the initial stage where there are no attributes, the hint view should ask the easiest question. Specifically, we calculate the pair-wise distances of visual exemplars in the mutual mental space and rank these pairs based on the semantic distances. Then we choose the most distant classes A and B in the embedding semantic space and ask the analyst \textit{what is the most visible attribute between classes A and B}.

\begin{figure}
 \centering 
 \includegraphics[width=1.0\linewidth]{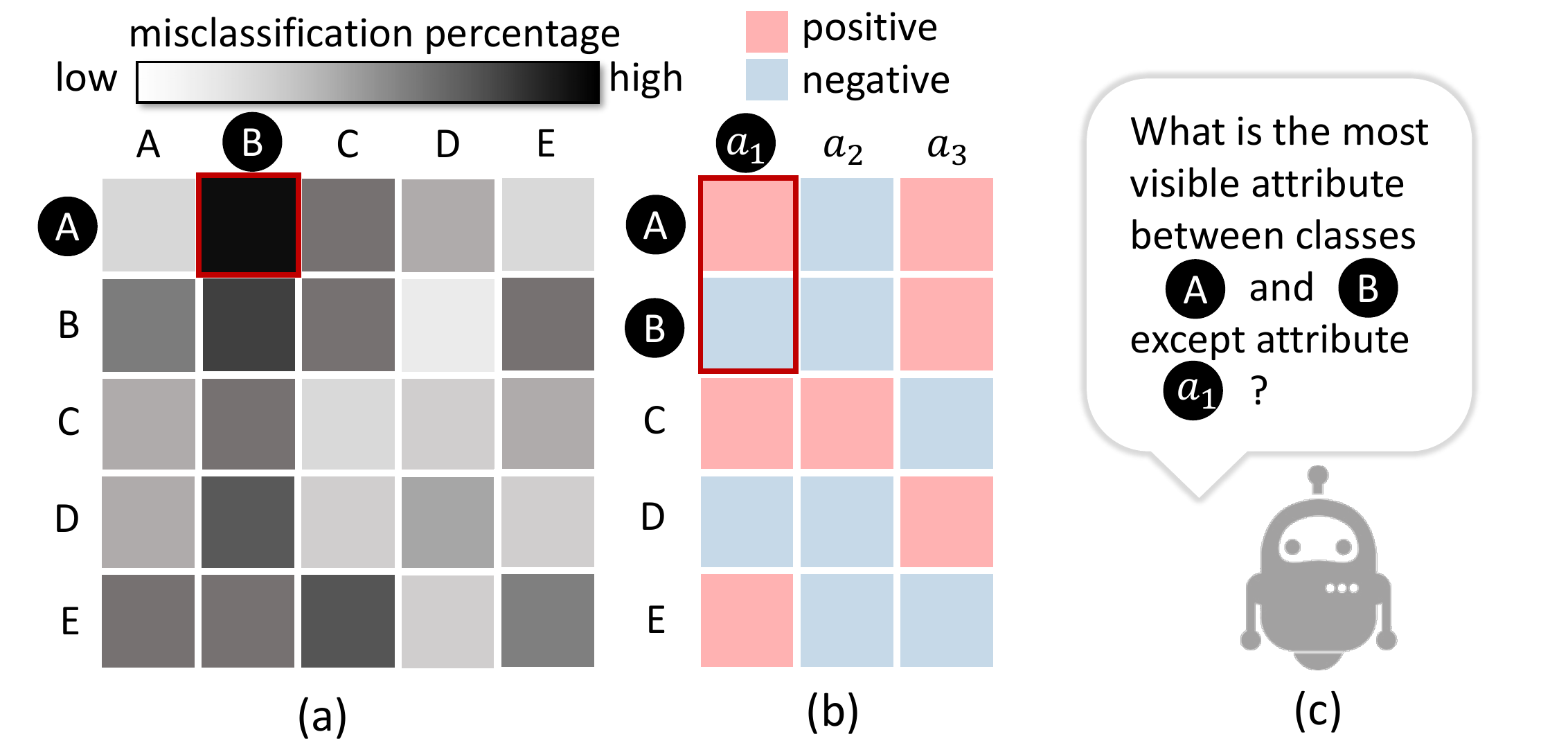}
 \caption{An example of generating questions and label recommendations. (a) A confusion matrix whose each cell is the percentage of misclassified instances in the training dataset. Classes A and B are most confused by the zero-shot classifier. (b) The class-attribute matrix of classes A-E with attributes $a_1$-$a_3$. Only attribute $a_1$ differentiates classes A and B. (c) The question generated in this situation is \textit{what is the most visible attribute between classes A and B except attribute $a_1$}.}
 \label{fig:hints}
\end{figure}


\begin{figure}
 \centering 
 \includegraphics[width=1.0\linewidth]{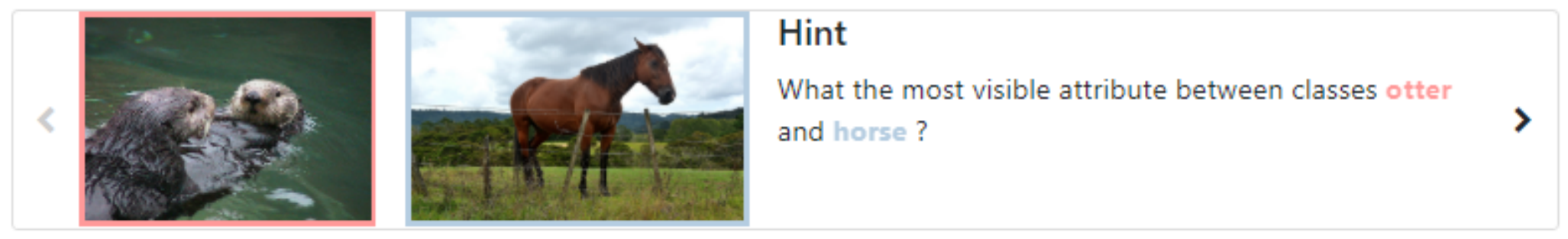}
 \caption{One contrastive question shown in the hint view. }
 \label{fig:hint-view}
\end{figure}

When the class-attribute matrix is not empty, the semantic navigator first builds a confusion matrix after retraining the zero-shot classifier. Each cell of the confusion matrix is the percentage of misclassification for that class in the training data. We use the percentage rather than the number to eliminate the influence of imbalanced data. When there are some classes with identical prototypes, we need to come up with a new attribute to differentiate these classes. Therefore, the semantic navigator selects a pair of classes that the zero-shot classifier confuses most from these classes. Suppose the classifier confuses classes A and B most, then the semantic navigator asks analysts \textit{what is the most visible attribute between classes A and B}. When all categories have different prototypes, the semantic navigator selects a pair of classes (A and B in Fig. \ref{fig:hints}(a)) that the zero-shot classifier confuses most. We then refer to the existing class-attribute matrix to find which attributes differentiate classes A and B. Suppose there are three attributes $a_1$, $a_2$ and $a_3$. Only attribute $a_1$ differentiates classes A and B while attributes $a_2$ and $a_3$ of classes A and B are the same (see Fig. \ref{fig:hints}(b)). The question is provided by asking \textit{what is the most visible attribute between classes A and B except attribute $a_1$}. The hint view ranks the questions by the percentage of misclassifications in the class-attribute matrix.


\begin{figure}
 \centering 
 \includegraphics[width=1.0\linewidth]{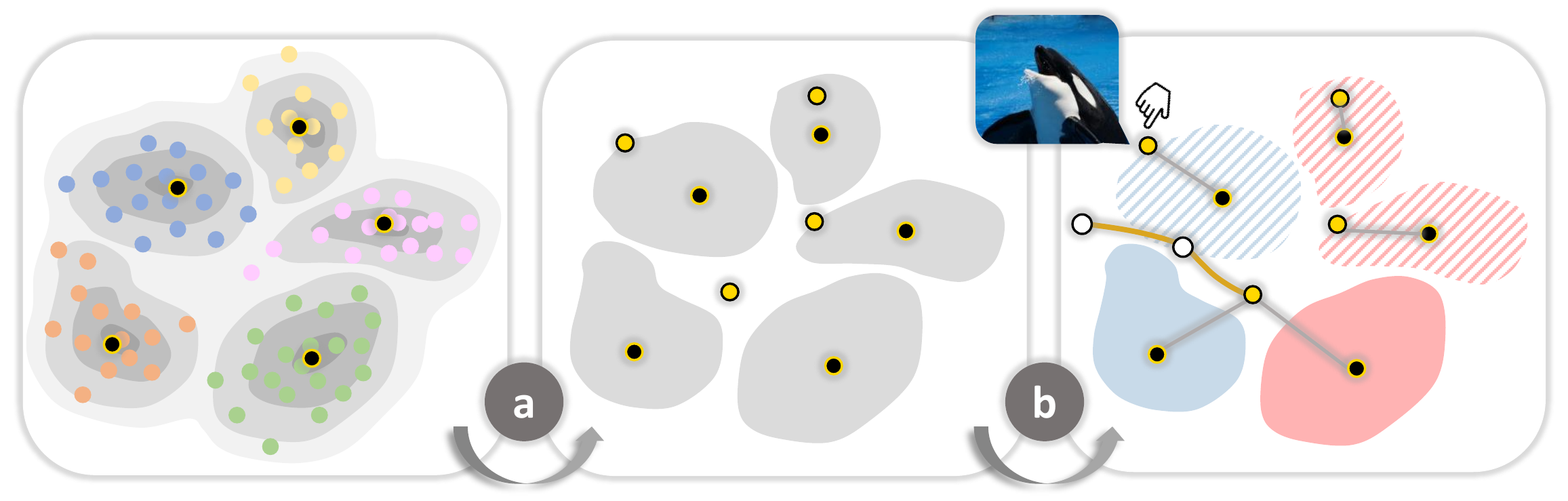}
 \caption{A conceptual pipeline of constructing the semantic map. (a) The semantic map abstracts the scatterplot as class-level contours. Meanwhile, it reprojects the prototypes every time the analyst adds a new attribute. (b) The semantic map enhances the visualization with exemplar-prototype links, nearest neighbors and semantic trajectories. }
 \label{fig:map-pipeline}
\end{figure}

\subsection{Visualization} \label{layout}

\noindent\textbf{Design Rationale.}
We visualize the mutual mental space by projecting the dataset $ \left\{M_{PCA}x_i^{tr}\right\} \cup V_s \cup \psi(P_s) $ into a 2D plane with two reasons. Firstly, the cluster structure is important to inspire analysts to define attributes since classes with similar semantics will cluster closely. Therefore, adjacent classes usually have the same attributes. Secondly, the dimension reduction can reveal the alignment of the class prototypes and the exemplars. As discussed in Sec. \ref{p:training}, the EXEM model learns a mapping function from the class prototypes to exemplars. Therefore, if class prototypes align well with the exemplars, the designed attributes have a good description of classes. We choose the t-SNE \cite{dermaaten2008visualizing} as the basic algorithm, as this is a popular nonlinear visualization technique to reveal the cluster structure.

The pipeline of constructing the semantic map can be viewed in Fig.\ref{fig:map-pipeline}. The semantic map abstracts the scatterplot as class-level contours and then projects the prototypes every time the analyst appends a new attribute. Finally, the semantic map enhances the visualization to support case-based reasoning and interactive feedback.

\noindent\textbf{Visual Abstraction.}
The semantic map abstracts the data distribution of each class as a contour by thresholding its density. Class exemplars lie inside the contours with their class labels annotated (Fig.\ref{fig:map-pipeline}). This design overcomes the overplotting problem and cognition load if all data are plotted, as we focus on the class-level semantics rather than observation-level semantics. Besides, we do not use colors to distinguish different classes for scalability issue, because zero-shot learning usually needs many classes to learn the transferable semantics. For example, the AWA2 dataset \cite{xian2019zero-shot}, a benchmark with minimum classes for zero-shot learning, already has forty classes for training and ten classes for testing.

\noindent\textbf{Reprojection.}
Every time the analyst appends a new attribute, the semantic map needs to update the prototypes' coordinates and anchor other data points, since only prototypes change during this process, and maintaining cognitive continuity is essential. One possible method is training a surrogate model such as a neural network to map from the PCA projection to the 2D t-SNE projection \cite{van2009learning}. Then we can only update the positions of class prototypes in every iteration. However, we do not adopt this method because learning a good approximation needs to design a good neural network architecture by empirically choosing different hyperparameters such as the number of layers and neurons. Instead we adopt another strategy by fixing the coordinates of $\left\{M_{PCA}x_i^{tr}\right\} \cup V_s$ and optimizing t-SNE loss function with stochastic gradient descent algorithm to update the positions of class prototypes $\psi(P_s)$ in every iteration. This method enables faithful projection and maintains cognitive continuity with fast computation.

\noindent\textbf{Visual Enhancement.}
The semantic map links the class exemplar and its corresponding prototype. This simple design enables case-based reasoning using exemplars and prototypes by comparing their relationships (details in Sec. \ref{pattern}). As exemplars and prototypes do not correspond to specific images, we enable the analyst to check their nearest neighbors to estimate their semantics. By examining the nearest images around class exemplars and prototypes, analysts can understand their semantic difference and leverage their inference ability to develop a new attribute hypothesis. Besides, the prototypes will travel around the semantic map and leave different trajectories during the design process. As we will show in Sec. \ref{use_cases}, the trajectories enable analysts to track the conditions of the model and understand how different attributes influence the model. For example, some prototypes will quickly converge to the visual exemplars after several iterations, which means the designed attributes describe these classes well. However, some prototypes will jump around two or three exemplars before the final convergence, which may infer that these prototypes confuse these classes easily. Therefore, we enable analysts to click class exemplars and show the trajectories of history positions. We name these trajectories as \textit{semantic trajectories}.

\subsection{Explanation with the Semantic Map} \label{pattern}

\begingroup
\setlength{\columnsep}{3pt}%
\setlength{\intextsep}{3pt}
\begin{wrapfigure}{l}{0.1\linewidth} 
    \centering
    \includegraphics[width=\linewidth]{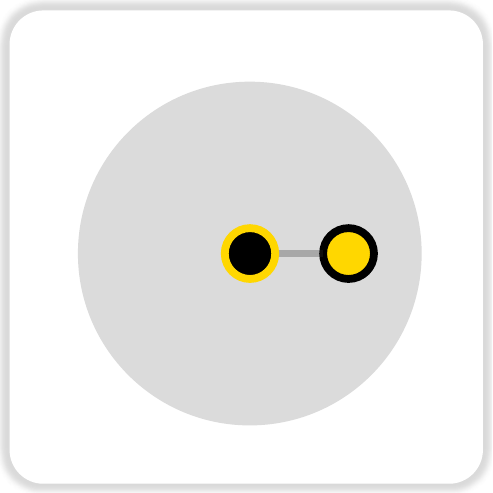}
\end{wrapfigure}

Now we present how semantic map supports explanations. As discussed above, each contour contains most of the data of that class. Therefore, if the class prototype comes inside the contour, it means the prototype has a high probability of having similar semantics with the exemplar inside the contour (see examples in Fig. \ref{fig:teaser}(a) for classes bobcat, fox, and wolf). We can sanity-check this assumption by examining the nearest neighbor images of exemplars and prototypes. This visual clue is the basic reasoning rule in our design.

\endgroup

When the visual exemplar comes outside the contour, we need to reason why the visual exemplar is not well aligned with its corresponding prototype. Reasoning this requires analysts to consider the context around this class and how class prototypes, contours, and exemplars interact with each other. Consider two classes A and B as the primitive case since it can be extended to cases involved multiple classes. Denote the contour of A and the class prototype of A as $C(A)$ and $P(A)$, respectively. The symbolism of B is the same. $P(A) \in C(A)$ stands for the class prototype lies inside the contour. We first discuss the situation where contours A and B do not intersect. In terms of whether $P(A) = P(B)$ and class prototypes inside the contours or not, there are five different visual patterns except the symmetrical cases.

\begingroup
\setlength{\columnsep}{3pt}%
\setlength{\intextsep}{3pt}
\begin{wrapfigure}{l}{0.2\linewidth} 
    \centering
    \includegraphics[width=\linewidth]{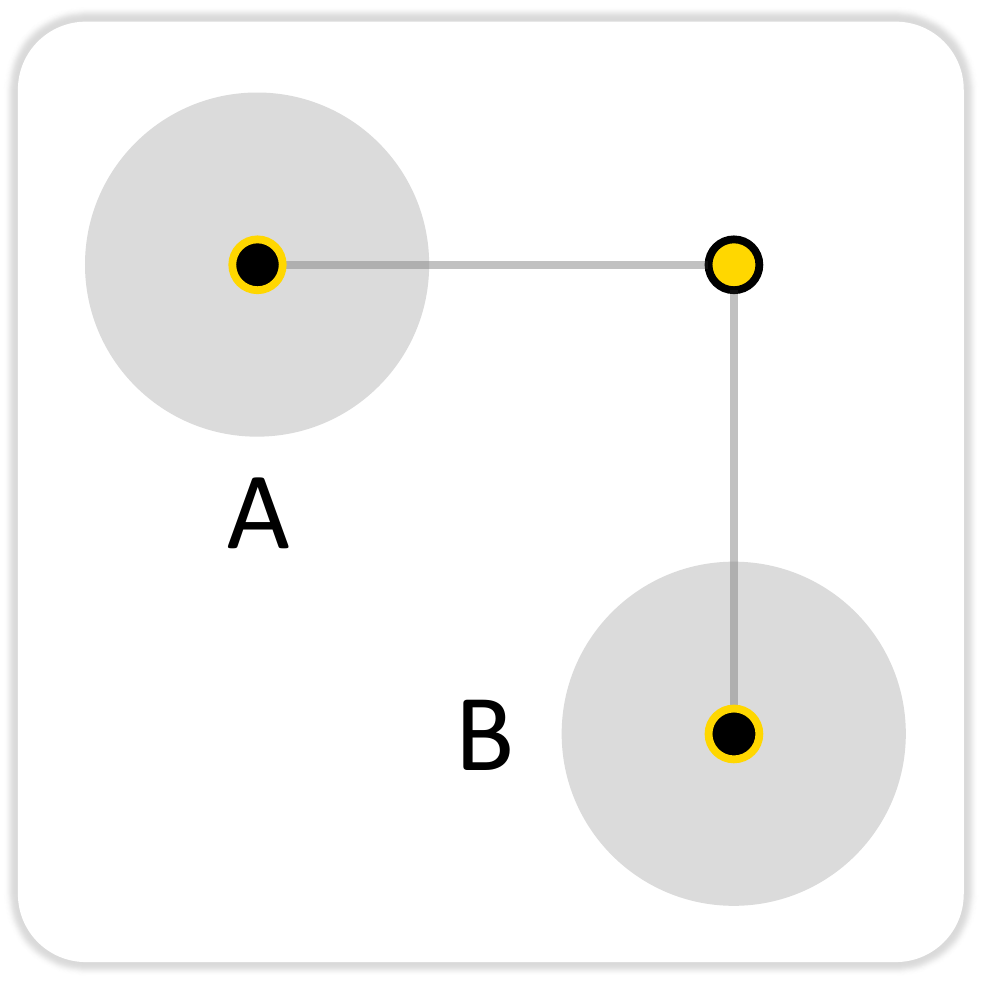}
\end{wrapfigure}

\noindent \textbf{Pattern 1.} $P(A) \notin C(A)$, $P(B) \notin C(B)$, $P(A) \notin C(B)$, $P(B) \notin C(A)$ and $P(A) = P(B)$ (or exchange A and B). Leveraging projection technique in Sec. \ref{layout}, the same data points will be projected at the same place. When class prototypes collapse into one point, it means classes A and B have the same attributes. Since both $P(A)$ and $P(B)$ are not inside the corresponding contours, we can infer that there is still a large gap between the prototypes and exemplars. Facing this situation, analysts should compare classes A and B to find the most distinguishable attribute that differentiates them.

\endgroup

\begingroup
\setlength{\columnsep}{3pt}%
\setlength{\intextsep}{3pt}
\begin{wrapfigure}{l}{0.2\linewidth} 
    \centering
    \includegraphics[width=\linewidth]{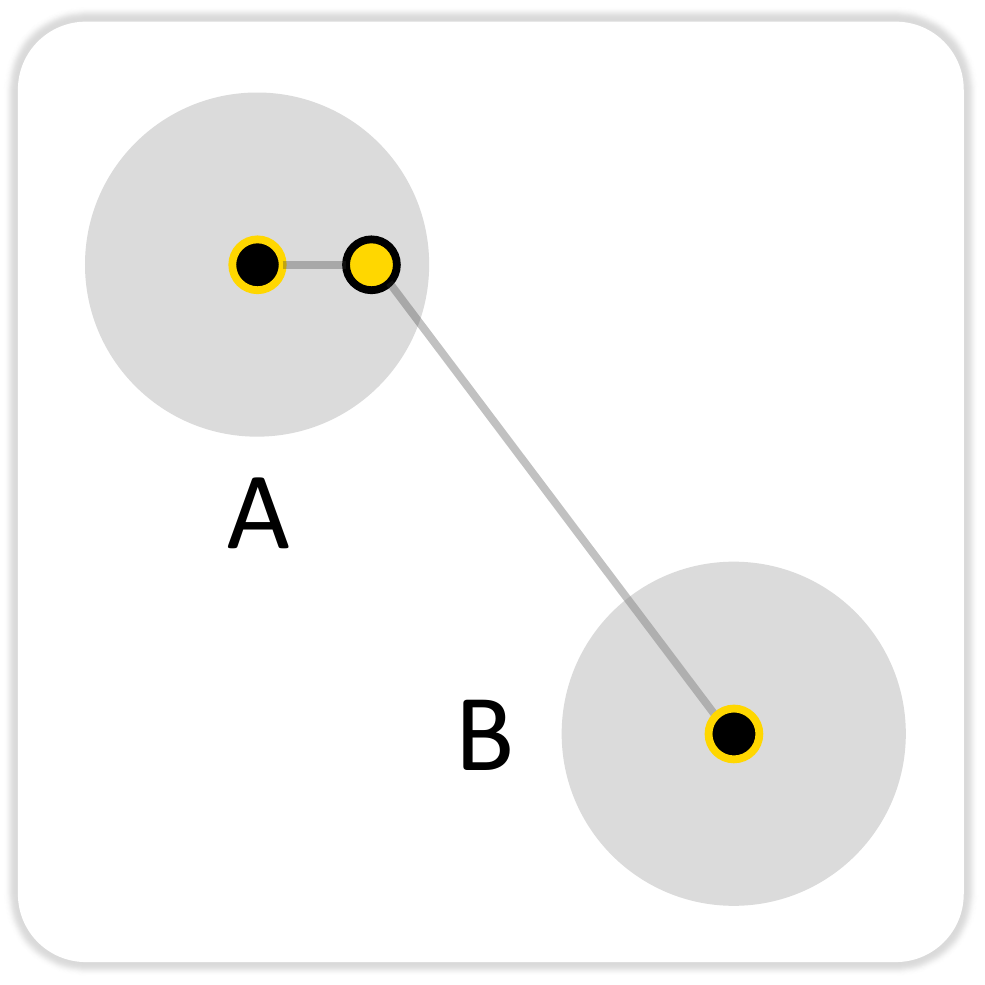}
\end{wrapfigure}

\noindent \textbf{Pattern 2.} $P(A) \in C(A)$, $P(B) \in C(A)$ and $P(A) = P(B)$ (or exchange A and B). This pattern shows that class prototype A aligns well with exemplar A. In contrast, prototype B does not align well with exemplar B. Since prototype A and B have the same attributes, analysts should adopt the same strategy as that of pattern 1 to figure out at least one attribute that differentiates A from B. Notice that pattern 1 and pattern 2 frequently happen at the early stage of the design process. When multiple classes are involved, these patterns will result in a star graph (see the subgraph circled by the dashed line in Fig. \ref{fig:case1}(e) as an example).

\endgroup

\begingroup
\setlength{\columnsep}{3pt}%
\setlength{\intextsep}{3pt}
\begin{wrapfigure}{l}{0.2\linewidth} 
    \centering
    \includegraphics[width=\linewidth]{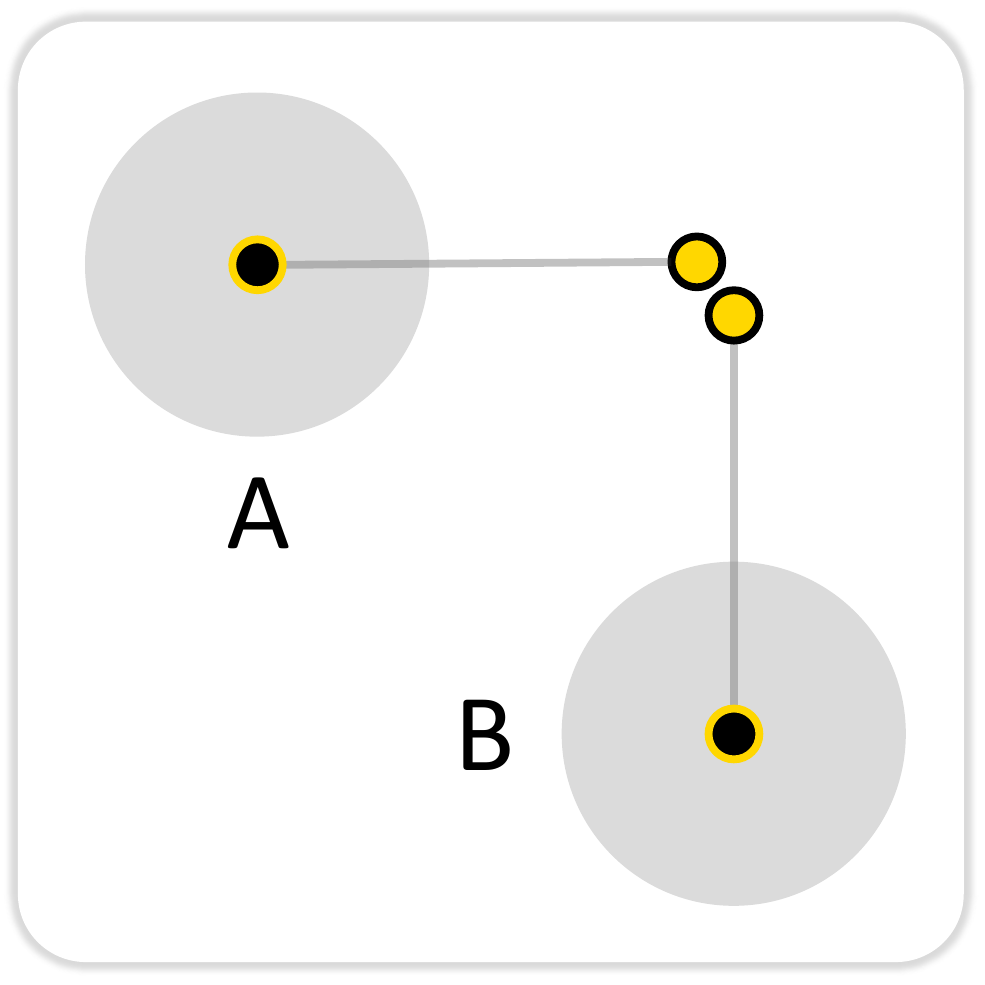}
\end{wrapfigure}

\noindent \textbf{Pattern 3.} $P(A) \notin C(A)$, $P(B) \notin C(B)$, $P(A) \notin C(B)$, $P(B) \notin C(A)$ and $P(A) \ne P(B)$ (or exchange A and B). Now analysts find at least one attribute that distinguishes class A and class B. However, the attributes do not describe both classes well. The reason for this pattern is the incompleteness of the attributes. To solve this problem, analysts should closely examine existing attributes and the corresponding exemplar images and check their missing attributes.

\endgroup

\begingroup
\setlength{\columnsep}{3pt}%
\setlength{\intextsep}{3pt}
\begin{wrapfigure}{l}{0.2\linewidth} 
    \centering
    \includegraphics[width=\linewidth]{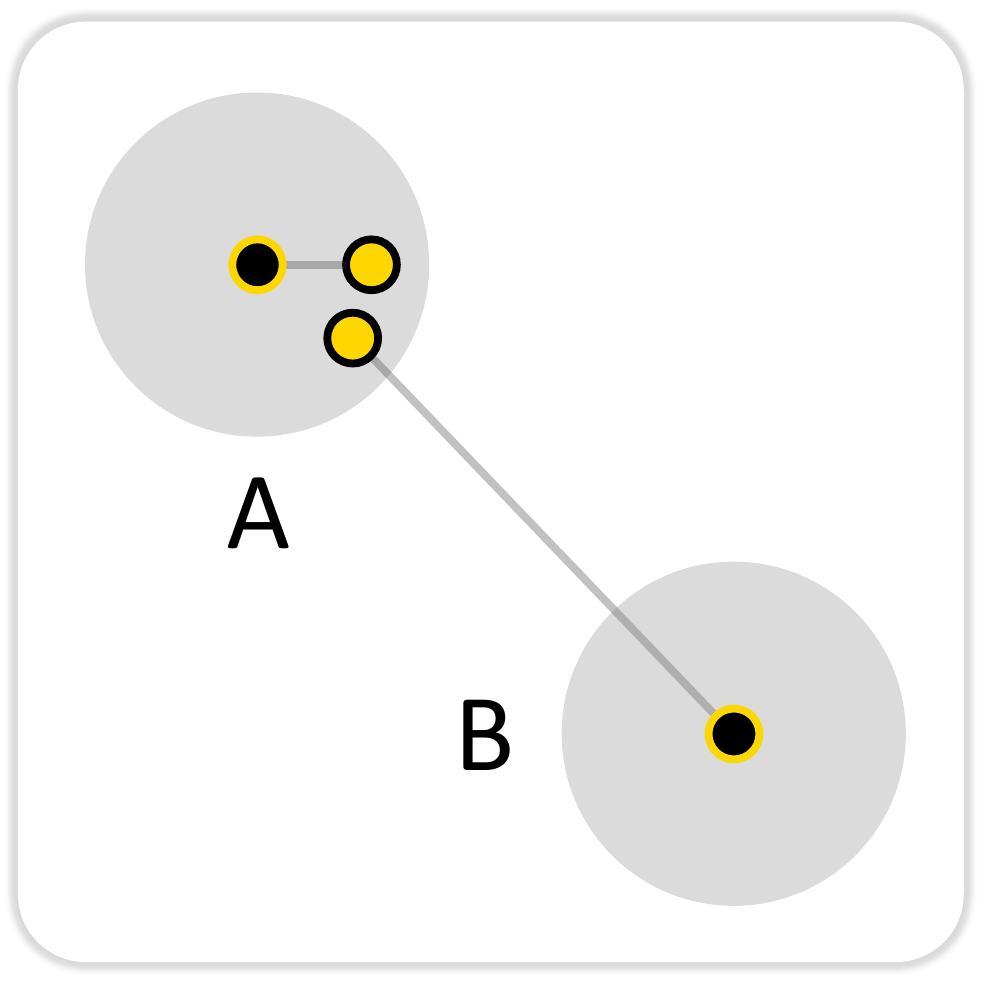}
\end{wrapfigure}

\noindent \textbf{Pattern 4.} $P(A) \in C(A)$, $P(B) \in C(A)$ and $P(A) \ne P(B)$ (or exchange A and B). The difference between this pattern and pattern 2 is that the two prototypes are different. However, prototype B is still similar to class A. It means the current attributes of class B are not well described. Therefore, analysts should further explore more attributes that distinguish B from other classes.

\endgroup

\begingroup
\setlength{\columnsep}{3pt}%
\setlength{\intextsep}{3pt}
\begin{wrapfigure}{l}{0.2\linewidth} 
    \centering
    \includegraphics[width=\linewidth]{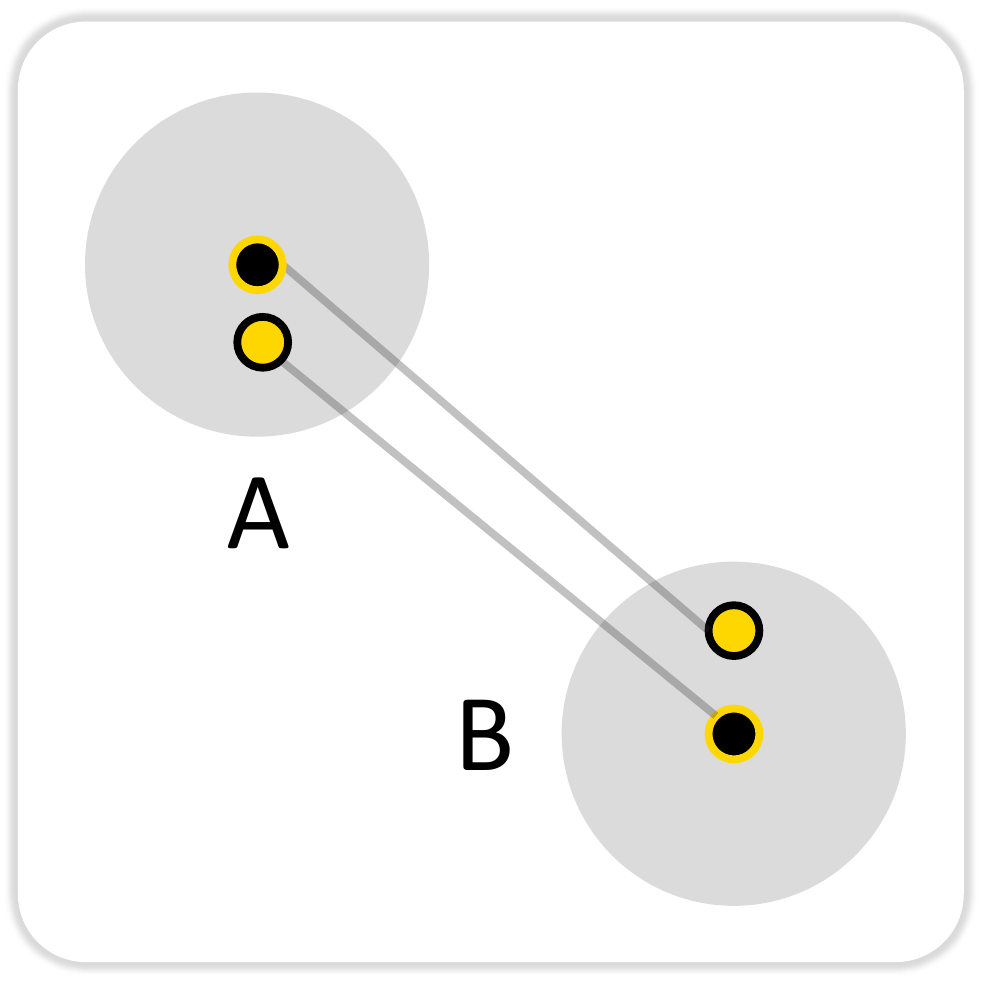}
\end{wrapfigure}

\noindent \textbf{Pattern 5.} $P(A) \in C(B)$ and $P(B) \in C(A)$ (or exchange A and B). This pattern shows that the machine wrongly recognizes A as B, and B as A. The reason behind this pattern may come from the errors in the class-attribute matrix. Therefore, analysts should closely examine the attributes to see whether they describe the wrong things. Hopefully, this problem can be solved by debugging the matrix.

\endgroup

The above five patterns consider the conditions in which $C(A)$ and $C(B)$ do not overlap. However, when $C(A)$ and $C(B)$ mostly intersect, the situation becomes complex.  Since most data points of classes A and B mix together in the feature space, it will be difficult to distinguish the two classes even though the attributes are distinguishable. In this situation, it will be better to improve the features first.


Now we discuss how the machine generates contrastive questions and label recommendations, and how analysts provide feedback to the recommendations through the interaction with the semantic map.

\subsection{Label Recommendation} \label{att_design}

Besides providing questions, the semantic navigator highlights the contours of selected classes and provides its attribute label recommendations by coloring class contours using red or blue striped textures. If this specification mismatches with the analyst's hypothesis, he/she can provide feedback to the machine with interactions discussed in Sec. \ref{interaction}. Then the machine adjusts the hypothesis correspondingly.

There are many hypotheses given the constrain that some classes are positive and some negative. The machine should provide the most distinguishable attribute hypothesis. Therefore, we leverage the semi-supervised support vector machine ($S^3VM$) \cite{bennett1999semi} to do this. Given the training set of labeled data and a working set of unlabeled data, $S^3VM$ searches for the best labeling of the working set that results in a support vector machine with maximum margin.

However, solving the $S^3VM$ problem is very challenging for its non-convexity and computational cost \cite{geng2019scalable}. It is still an open problem to scale up $S^3VM$ for large-scale applications. Using the whole dataset as the training and working set is unrealistic because analysts need instant feedback when generating attribute hypotheses. Considering this, we build $S^3VM$ using class exemplars rather than the whole datasets as they have already summarized the cluster structure. We solve this problem using quasi-Newton schemes, the state-of-art method namely $QN$-$S^3VM$ \cite{gieseke2014fast}.

\begin{figure*}
 \centering 
 \includegraphics[width=1.0\textwidth]{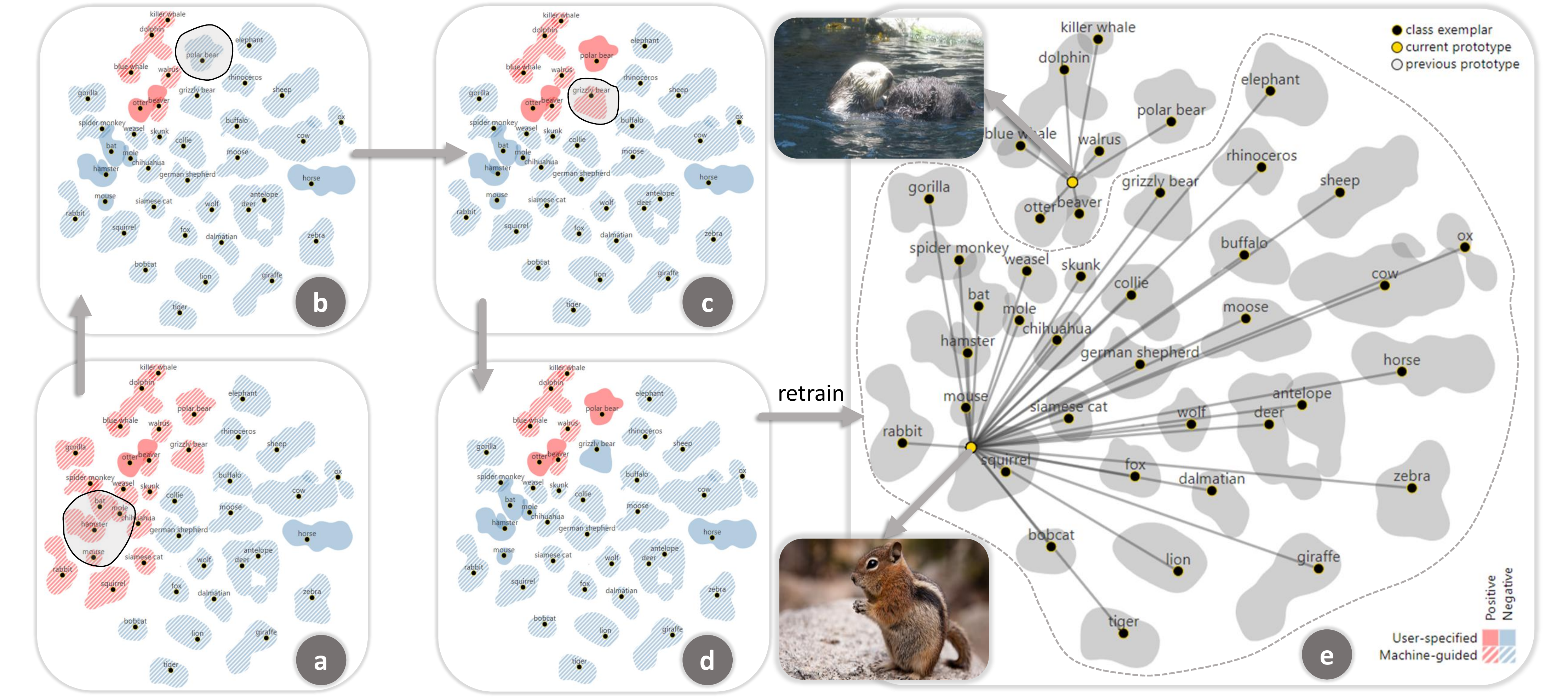}
 \caption{The design process of water attribute: (a) The analyst annotates hamster, bat, mole, and mouse as negative using a lasso to modify the attribute specification recommended by the machine. (b) The machine adjusts the attribute specification. The analyst circles the polar bear and labels it as positive. (c) The result is updated by the machine. The analyst labels the grizzly bear as negative. (d) The machine agrees with the analyst and does not change its decision. (e) The result after retraining the model using the water attribute. }
 \label{fig:case1}
\end{figure*}

\begingroup
\setlength{\columnsep}{3pt}%
\setlength{\intextsep}{3pt}
\begin{wrapfigure}{l}{0.3\linewidth} 
    \centering
    \includegraphics[width=\linewidth]{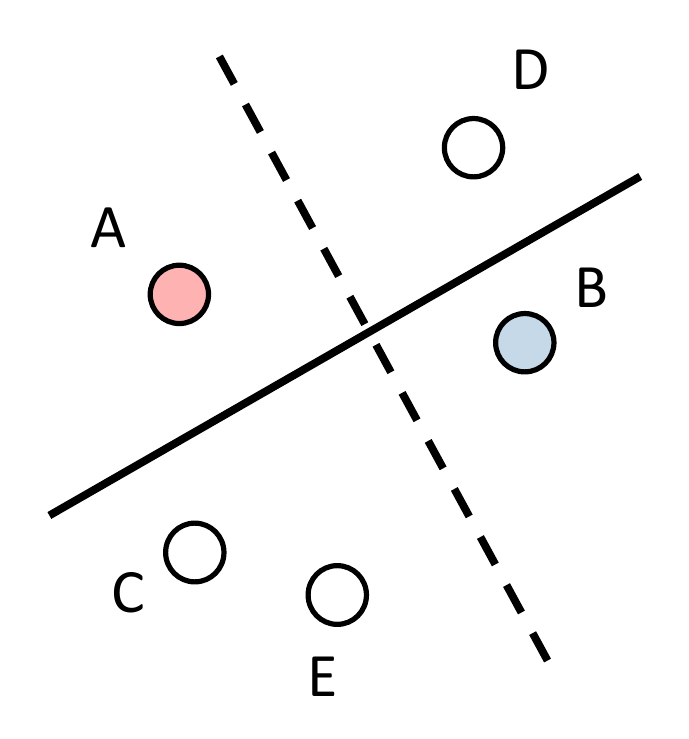}
\end{wrapfigure}

For example, suppose the analyst adopts the hint that asks him/her to come up with a new attribute to distinguish classes A and B except for the attributes $\left\{a_i\right\}_{i=1}^n$ (see one example in Fig. \ref{fig:hints} (c)). Note that $\left\{a_i\right\}_{i=1}^n$ are the existing attributes that can distinguish classes A and B. Therefore, the question asks for a new attribute to distinguish the two classes except for $\left\{a_i\right\}_{i=1}^n$. Otherwise, the analyst may come up with a redundant attribute specified before. Then, we use $QN$-$S^3VM$ to find a hyperplane in the mutual mental space that results in the maximum margin with class A as positive and class B as negative. The hyperplane splits the mutual mental space into two sub-spaces. All the other classes lie on the same side as class A will be labeled as positive, and the other classes labeled as negative. The semantic navigator enables analysts to reverse the status of the two groups of classes (change positive classes as negative and vice versa) using this button \faRetweet. If this hyperplane happens to result in the attribute $a_i$, we choose the hyperplane with the next maximum margin until there is no conflict.

\endgroup


\subsection{User Feedback} \label{interaction}


Analyzing the visual patterns in the semantic map guided by questions, analysts can generate an attribute hypothesis and specify it by correcting label recommendations. Specifically, the analyst can click the class contours to choose whether they are positive or negative. We use red for positive and blue for negative. Since classes with similar semantic locate closely, they usually have the same attributes. Therefore, we enable analysts to select multiple classes by drawing lassoes.  The analyst can modify the label recommendations by interactively specifying partial classes. And the machine predicts the labels of the left classes at the same time. This collaboration eases the labeling burden and improves the efficiency of humans.

After generating an attribute hypothesis, the analyst finetunes the specification in the matrix view, names the attribute hypothesis, and clicks the submit button \faArrowRight, which triggers a popup window to let him/her choose images for unseen classes with the newly added attribute. These images are the nearest neighbors for each unseen class exemplar. We show the images to ease the recall of the semantics of unseen classes for the analyst. However, it is not required as there may be no images for unseen classes in an extreme situation. In this situation, the images are replaced by their names. Once committed, the underlying EXEM model will be retrained automatically. Meanwhile, the visualization will be updated as well. A new iteration begins until the stopping condition is met.


\section{Evaluation}

We conduct case studies and controlled user studies to evaluate our method.



\subsection{Case Studies} \label{use_cases}

We first demonstrate the effectiveness of the semantic navigator through case studies using the Animals with Attributes2 (AWA2) dataset \cite{xian2019zero-shot} \footnote{AWA2 and its images, labels, attributes, and features are publicly available in https://cvml.ist.ac.at/AwA2/}, a standard benchmark for zero-shot classification. AWA2 has 37322 images of 50 kinds of mammals, along with the features extracted from a pre-trained ResNet \cite{he2016deep}. We use the standard split of the dataset (40 classes for training and 10 for testing) \cite{xian2019zero-shot} for the study. Besides, AWA2 provides a class-attribute matrix with 85 binary attributes. This class-attribute matrix is originally collected in a psychological experiment \cite{osherson1991default} on attribute-based object similarity from human subjects.





Suppose one data scientist Bob wants to build an animal classifier using the AWA2 dataset. He first explores the semantic map to get familiar with the animals. Then he finds that similar animals in the semantic map are close to each other, while different animals are far apart.


\begin{figure*}
 \centering 
 \includegraphics[width=1.0\textwidth]{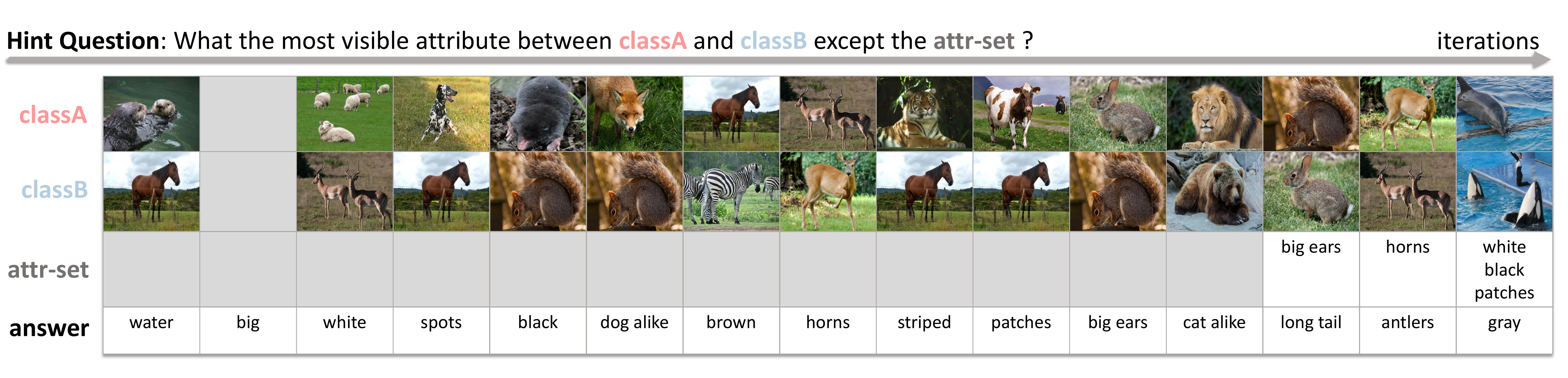}
 \caption{The hint questions asked by the machine during the design process of the class-attribute matrix, along with answers provided by the analyst in Sec. \ref{use_cases}. Notice that in the second iteration, the analyst comes up with the big attribute by himself rather than inspired by the machine. }
 \label{fig:directing_guidance}
\end{figure*}

\noindent\textbf{Semantic navigator orients the modification process of attributes.}
Bob checks the hint view and preview a list of recommendations. The machine first asks him \textit{what is the most visible attribute between classes otter and horse} (Fig. \ref{fig:hint-view}). Bob quickly notices that the otter lives in the water, while the horse lives on the ground. Therefore, he decides to add water concept to the empty class-attribute matrix. The semantic navigator provides an attribute hypothesis shown in the semantic map (Fig. \ref{fig:case1}(a)). The machine thinks this attribute distinguishes the otter and horse most. However, this attribute does not align with the water concept of what Bob thinks. Therefore, Bob decides to modify the attribute hypothesis. He first circles the bat, mole, hamster, and mouse, and labels these animals as negative (Fig. \ref{fig:case1}(a)). Then the machine quickly gives feedback to adjust the attribute hypothesis. The updated attribute hypothesis is shown in Fig. \ref{fig:case1}(b). Bob notices that the semantic navigator omits the polar bear, as the polar bear usually jumps into the sea to catch fish. Therefore, Bob circles the polar bear and marks it as positive as well (Fig. \ref{fig:case1}(b)). Then the machine provides feedback and recommends labeling the grizzly bear as positive (Fig. \ref{fig:case1}(c)). It seems that the machine has a different opinion of the attribute hypothesis compared with that of Bob. However, Bob is quite sure that grizzly bear usually lives in the forest. Therefore, he modifies the hypothesis and labels the grizzly bear as negative (Fig. \ref{fig:case1}(c)). Now, the machine seems to agree with Bob and does not change the hypothesis anymore (Fig. \ref{fig:case1}(d)). Bob examines the specification closely, then names the hypothesis as water and summit it. Then the semantic navigator pops up a window to ask Bob to specify the unseen classes by choosing images with the attribute water. There are ten kinds of unseen animals, including chimpanzee, giant panda, leopard, persian cat, pig, hippopotamus, humpback whale, raccoon, rat, and seal. This task is effortless, and Bob quickly selects the hippopotamus, humpback whale, and seal. Finally, Bob commits the task and retrains the underlying zero-shot model with the water attribute. The final result is shown in Fig. \ref{fig:case1}(e).


\noindent\textbf{Semantic navigator directs the design process of the class-attribute matrix step by step.}
In the following iterations, Bob is guided by the machine by answering the questions in the hint view. Questions and corresponding answers of the first fifteen attributes specified by Bob are shown in Fig. \ref{fig:directing_guidance}. It seems that it is getting more difficult to answer the questions when the iteration increases. For example, the machine asks Bob what is the most visible attribute between dolphin and killer whale except the white, black, and patches attributes in the final iteration. Nevertheless, with the collaboration between the machine and Bob, semantic prototypes are getting closer to their corresponding class exemplars (Fig. \ref{fig:teaser}(a)). For example, in Fig. \ref{fig:teaser}(a), the semantic trajectory of the sheep presents that the sheep prototype transforms from squirrel and buffalo to sheep and polar bear, and finally gets to the sheep. Most of the semantic prototypes align well with their corresponding class exemplars such as siamese cat, fox, and wolf. Guided by the semantic navigator, Bob creates a better class-attribute matrix with only 15 visual attributes without losing the interpretability. This new class-attribute results in comparable training accuracy (ours: 79.612\%, baseline: 89.076\%) and higher testing accuracy (ours: 70.966\%, baseline: 57.079\%) compared with the baseline (EXEM \cite{changpinyo2017predicting}) with 85 attributes (see Fig. \ref{fig:teaser}(d)).


\subsection{User Studies} \label{user_studies}

We conduct user studies to evaluate the semantic navigator \footnote{The semantic navigator is publicly available at \href{https://bit.ly/3y4yZQl}{https://bit.ly/3y4yZQl}.} using both AWA2 and a subset of CUB200-2011 \cite{wah2011caltech} \footnote{CUB200-2011 and its images, labels, and attributes are publicly available in http://www.vision.caltech.edu/visipedia/CUB-200-2011.html}. CUB200-2011 is a fine-grained bird dataset containing 200 bird species and binary 312 attributes with 6033 images. These attributes are collected from MTurk workers \cite{wah2011caltech} guided by 28 attribute questions based on an online tool for bird species identification \footnote{http://www.whatbird.com/}. The standard split of this dataset contains 150 training categories and 50 testing categories. We select a subset of the CUB-200-2011 for evaluation by randomly selecting 40 training classes and ten testing classes from the original training classes and testing classes, respectively. We name the selected dataset as SUB-CUB200 for short. The features for both AWA2 and SUB-CUB200 are 2048-dimensional extracted from the last pooling layer of ResNet \cite{he2016deep}.

Twenty-six participants (denote as P1-P26) in a large university volunteer the studies, including nine undergraduates, thirteen masters, and four Ph.D. students. To compare with our system, we design a system without guidance. This system contains only two views, including a matrix view (Fig. \ref{fig:teaser}(f)) and a line chart (Fig. \ref{fig:teaser}(e)) as the semantic navigator. It enables analysts to append a new column and modify cells to specify a new attribute. Whenever the analyst adds a new attribute, the underlying zero-shot model is trained online and the line chart for accuracy is updated.

\begingroup
\setlength{\columnsep}{3pt}%
\setlength{\intextsep}{3pt}
\begin{wrapfigure}{l}{0.5\linewidth} 
    \centering
    \includegraphics[width=\linewidth]{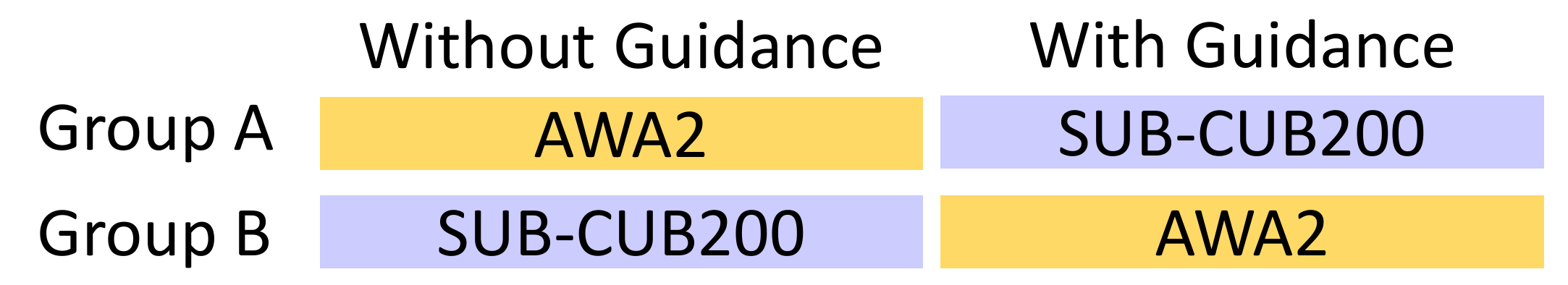}
\end{wrapfigure}

We divide the participants into two groups (group A and group B) randomly and evenly. Before all the experiments, participants should fill in a questionnaire to provide personal information. Both groups need to conduct two experiments without and with guidance.  Participants in group A build one classifier for AWA2 without guidance, while the other one for SUB-CUB200 with guidance. In contrast, group B reverses the datasets. We require participants to accomplish classification tasks with two different datasets to mitigate the anchoring bias if the same dataset is conducted sequentially. Besides, we conduct the two experiments on separate days in case the participants are exhausted.

\endgroup

Before each experiment, we conduct a workshop with each group to explain some functionality of the system without guidance or the semantic navigator and how to build a zero-shot classification model by describing animals using attributes. We do not explain any details of the underlying zero-shot model as participants can complete the task without understanding technical details. They only need to describe animals objectively with attributes that are visually distinguishable. Participants can explore the system and ask any questions during this session. This session lasts about thirty minutes. After each experiment, participants need to fill in a questionnaire to provide some feedback.

During each experiment, every participant has to design 15 attributes at the completion of the task. It takes each participant about 30-45 minutes to complete the task in each experiment. The whole design process is screen recorded. Notice that we adjust the system's language to their corresponding native language since the foreign words of animals can influence their cognition. We also allow analysts to name the attributes in their native language. Moreover, we translate the attributes later on. After each experiment, the class-attribute matrixes are downloaded locally for further analysis. At the end of the user studies, we interview each participant to get qualitative feedback. We summarize the results and our findings in the following.

\begin{figure}
 \centering 
 \includegraphics[width=1.0\linewidth]{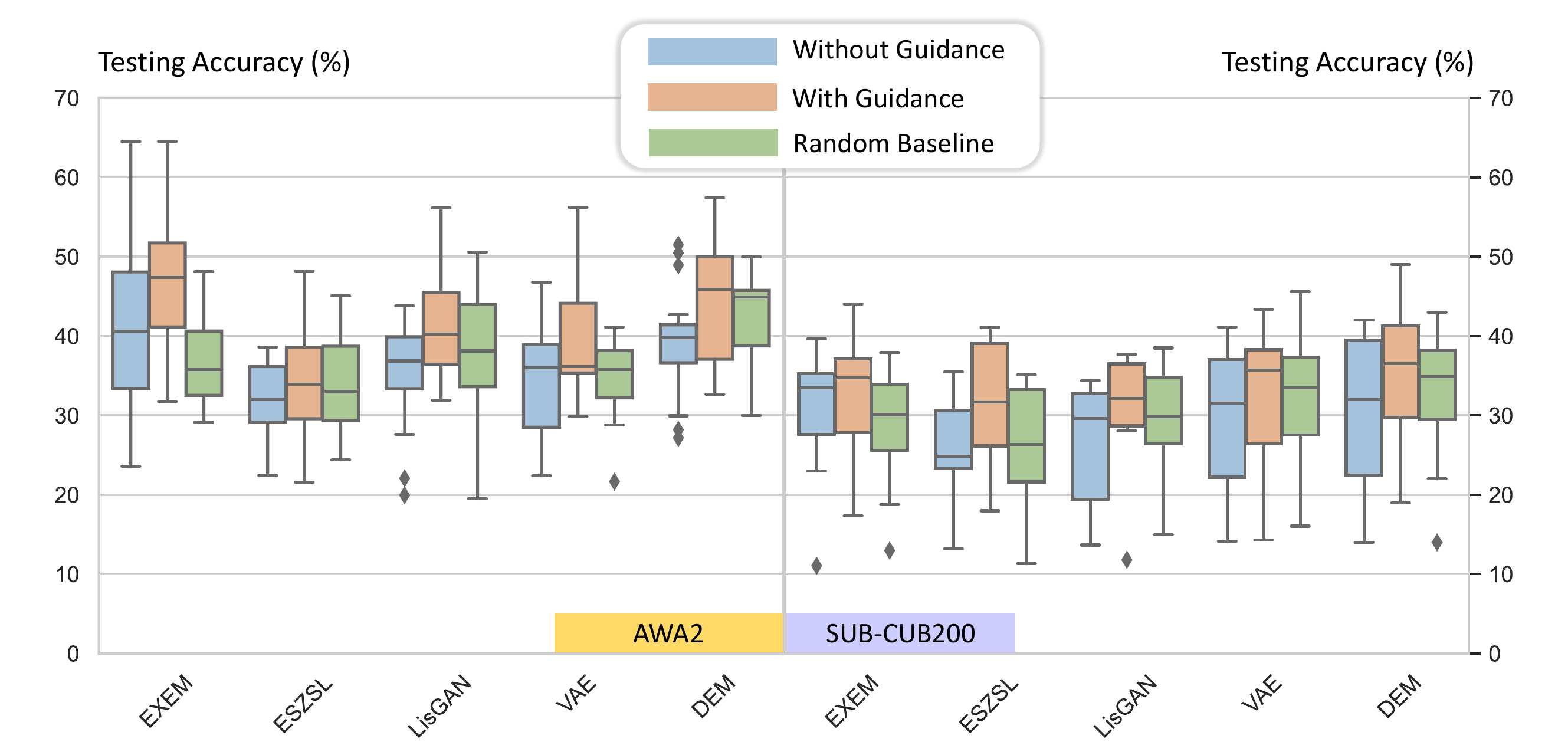}
 \caption{The box plot shows the testing accuracy of three settings (without guidance, with guidance, and random baseline) for two datasets (AWA2 and SUB-CUB200), evaluated by five state-of-art zero-shot algorithms (EXEM \cite{changpinyo2017predicting}, ESZSL \cite{romera2015embarrassingly}, LisGAN \cite{li2019leveraging}, VAE \cite{schonfeld2019generalized}, and DEM \cite{zhang2017learning}).  }
 \label{fig:testing_performance}
\end{figure}

\subsubsection{Quantitative results of zero-shot classification}

We compare the results from three settings by calculating the testing accuracy using several state-of-art zero-shot algorithms, including EXEM \cite{changpinyo2017predicting}, ESZSL \cite{romera2015embarrassingly}, LisGAN \cite{li2019leveraging}, VAE \cite{schonfeld2019generalized}, and DEM \cite{zhang2017learning} (Fig. \ref{fig:testing_performance}). Three settings include the method without guidance, method with guidance using the semantic navigator, and method using random strategy. The random strategy simulates 20 persons by randomly selecting 15 attributes from the class-attribute matrix benchmark. Besides testing accuracy, we also evaluate the results based on the average time to specify each attribute (total time divided by the number of attributes). The total time starts when the participant enters each experiment and ends after the retraining process and updating the semantic map with the final class-attribute matrix.



\begin{figure}
 \centering 
 \includegraphics[width=1.0\linewidth]{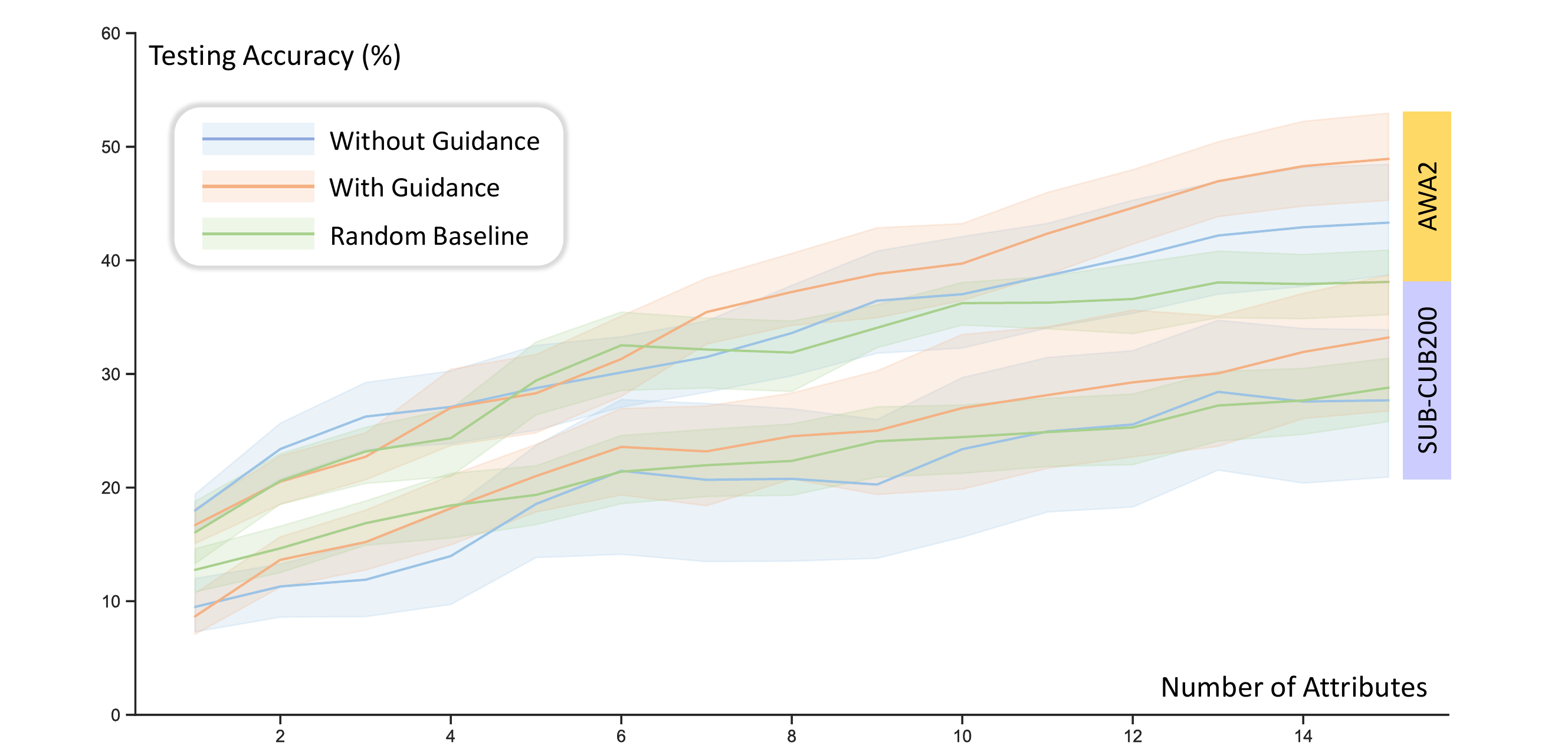}
 \caption{
The line chart compares the testing accuracy of two datasets under three settings (tested by EXEM algorithm). Each band summarizes the distribution of all corresponding performance curves. }
 \label{fig:comparison-result}
\end{figure}


\begingroup
\setlength{\columnsep}{3pt}%
\setlength{\intextsep}{3pt}
\begin{wrapfigure}{l}{0.5\linewidth} 
    \centering
    \includegraphics[width=\linewidth]{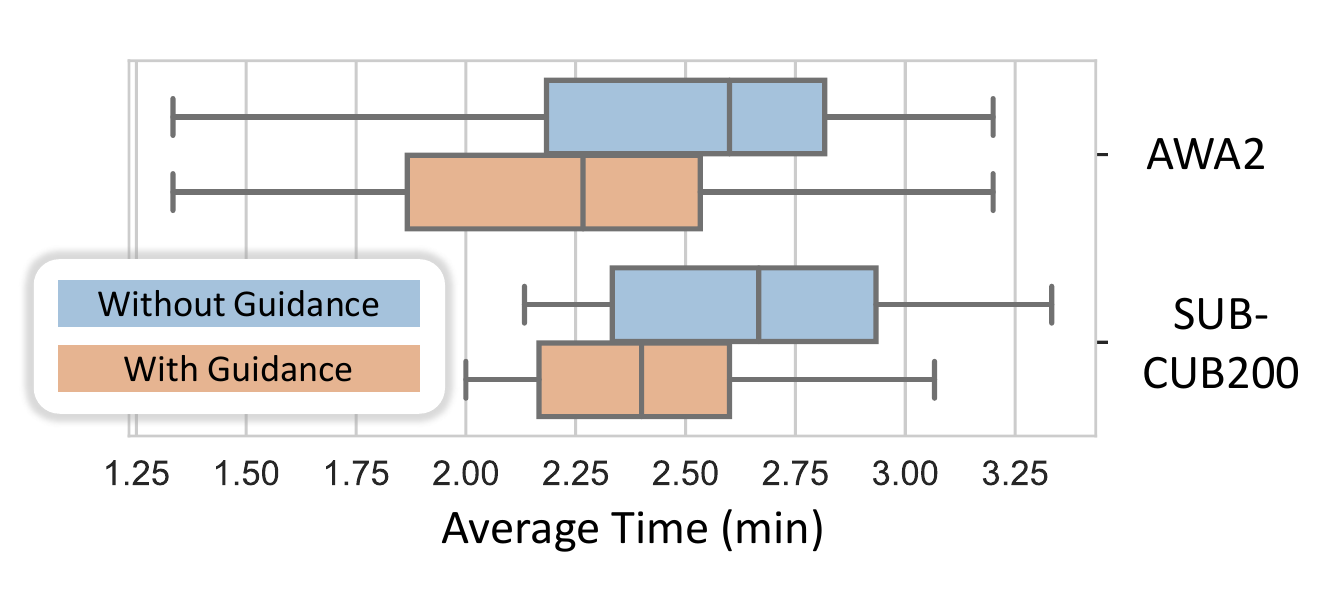}
    \caption{The box plot shows the average time to specify each attribute for each group of participants.}
    \label{fig:average-time}
\end{wrapfigure}

As shown in Fig. \ref{fig:average-time}, participants can spend less time coming up with new attributes for both datasets with the semantic navigator. Participants can save about 14.42 seconds to create one attribute for AWA2 and 14 seconds for SUB-CUB200 on average. As for testing accuracy (Fig. \ref{fig:testing_performance}), different algorithms produce various results in our studies. All three settings result in lower accuracy in SUB-CUB200 than AWA2 because the task of distinguishing birds in SUB-CUB200 is more challenging than the task of AWA2. Besides, participants spend more time completing the task of SUB-CUB200 than that of AWA2 regardless of whether with guidance. Moreover, participants without guidance gain comparable performance as the random baseline. However, we also find noticeable performance gain once analysts apply the semantic navigator. On average,
the semantic navigator can increase the test accuracy by 4.96\% for the AWA2 and 4.13\% for the SUB-CUB200. This difference can become notable when participants create over five attributes (Fig.\ref{fig:comparison-result}).

\endgroup


\subsubsection{Qualitative feedback}

More than half of the participants without guidance mentioned that it is hard for them to complete the task. Participant P2 said: \textit{ In the beginning, I think this task is effortless as what I have to do is just to find some words to describe animals. Everybody can do it. Nevertheless, I quickly found that it is not as easy as expected. Most time, when I specify a new attribute, the test accuracy does not change much and even decreases. It makes me crazy as I have no idea what's wrong I have done.} Participant P5 agrees with it and commented: \textit{At the end of the design process, I think my brain is stuck. I do not know what I have specified before. When I want to add a new attribute, I need to go back to check if I have specified it already. } Participant P8 also mentioned: \textit{It is very boring to create the matrix. When I come to a new attribute, I have to specify the animals individually to check whether they have the attribute. During the design process, I need to pay much focus and specify the matrix carefully. It exhausted me.}

However, most participants (18 / 26) with guidance thought that the semantic navigator is easy to understand and use. P11 mentioned that \textit{the questions helped me a lot. You just to follow its guidance to answer the questions only.} P23 commented that \textit{sometimes the questions are too hard for me. For example, what is the most visible attribute between the otter and the beaver? I know they are different, but I can not find an appropriate word to distinguish them. But the good thing is that you can skip the hard question and find one that you can answer.} P15 liked the design of the semantic map, he said \textit{at the beginning I did not understand what the black dots mean. But when you explained it as the destinations as the normal map, I quickly understand what I need to do is to let the yellow dots reach the corresponding destinations. I like it as it provides an overview, and I can understand the current status of the model. Meanwhile, the semantic map can tell me where I should pay attention.} One participant P17 acknowledged that the label recommendations are beneficial, he said: \textit{Whenever I fix the specification, the machine can recommend the rest. It impressed me. However, sometimes we disagreed with each other so that I need to specify the attributes with several interactions. It would be better if it could read my mind more efficiently. }


\section{Conclusion, Discussion and Future Work}

This work focuses on the fundamental problem of designing the class-attribute matrix for zero-shot classification with the mixed-initiative approach. We propose visual explainable active learning with four actions (ask, explain, recommend, and respond) to promote human-AI teaming. Besides, we design and implement a visual analytics system called semantic navigator for interactive zero-shot classification. To justify our method, we conduct case studies and controlled user studies. Results show that the semantic navigator improves analysts' efficiency for building zero-shot classifiers compared with the method without guidance.

Although the visual explainable active learning approach targets zero-shot classification in this paper, its concept has generalizability. The machine asks, and human answers using attributes can be viewed as a form of human-machine communication. At the same time, the visualization creates a shared space to bridge the low-level feature space and the high-level semantic space, which may help build the shared mental model between the human and machine. Our work contributes a new perspective on how humans and AI interact and collaborate via visual analytics and may inspire researchers to promote better human-AI teamwork. Future work can go beyond animal classification to real-world problems, such as the human-AI teaming approach to medical diagnosis \cite{cai2019human}.

In the future, we can further improve human-AI teaming with more effective human-AI communication and further explanations. Firstly, we start our work by considering only binary attributes. Future work can extend binary attributes to relative attributes \cite{parikh2011relative}. For example, the classifier can capture that animal A is furrier than animal B. Using relative attributes can foster more natural human-AI communication \cite{parikh2012relative} and more effective human feedback \cite{parkash2012attributes}. Secondly, we quantify the performance of zero-shot classifiers based on the model accuracy. However, one open question is how to know the classifier learns the attributes accurately rather than other related concepts? Recent explainable artificial intelligence research has shifted attention to quantifying concept-level explanations \cite{kim2018interpretability, ghorbani2019towards}, which can complement our work.

\acknowledgments{
This research is supported by the National Key Research and Development Program of China under Grant No.2019YFC1521200. }

\bibliographystyle{abbrv-doi}

\bibliography{template}
\end{document}